\documentclass[11pt]{article}
\pdfoutput=1
\usepackage{amsmath}
\usepackage{amsthm}
\usepackage{amscd}
\usepackage{amsfonts}
\usepackage[psamsfonts]{amssymb}
\usepackage{graphicx}
\usepackage{hyperref}
\usepackage{xcolor}

\numberwithin{equation}{section}

\makeatletter
\renewcommand\section{\@startsection {section}{1}{\z@}%
                                 {-3.5ex \@plus -1ex \@minus -.2ex}
                                   {2.3ex \@plus.2ex}%
                                   {\normalfont\large\bfseries}}
\renewcommand\subsection{\@startsection{subsection}{2}{\z@}%
                                   {-3.25ex\@plus -1ex \@minus -.2ex}%
                                     {1.5ex \@plus .2ex}%
                                     {\normalfont\bfseries}}
\renewcommand\subsubsection{\@startsection{subsubsection}{3}{\z@}%
                                   {-3.25ex\@plus -1ex \@minus -.2ex}%
                                     {1.5ex \@plus .2ex}%
                                     {\normalfont\itshape}}
\makeatother

\usepackage{slashed}

\def\+{\!+\!}
\def\-{\!-\!}
\def\={\!=\!}

\begin{document}

\begin{flushright}
\end{flushright}
\vspace{0.5in}

\begin{center}{\LARGE{\bf Entanglement Entropy with Background Gauge Fields  
} } 
\vspace{0.85in}

{\large \bf Bom Soo Kim}\\ ~\\ \vspace{-0.1in}
{\normalsize {Department of Physics and Astronomy,}} \\
{\normalsize {University of Kentucky, Lexington, KY 40506, USA}}\\ 
{\normalsize  \it  {bom.soo.kim@uky.edu}}\\ 
\end{center}
\bigskip 
\vspace{0.5in}

\begin{abstract}

We study the entanglement entropy, the R\'enyi entropy, and the mutual (R\'enyi) information of Dirac fermions on a 2 dimensional torus in the presence of constant gauge fields. We derive their general formulas using the equivalence between twisted boundary conditions and the background gauge fields. Novel and interesting physical consequences have been presented in arXiv:1705.01859. Here we provide detailed computations of the entropies and mutual information in a low temperature limit, a large radius limit, and a high temperature limit. The high temperature limit reveals rather different physical properties compared to those of the low temperature one: there exist two non-trivial limits that depend on a modulus parameter and are not smoothly connected. 

\end{abstract}

\newpage
\tableofcontents
\newpage 

\section{Introduction}

Quantum entanglement is a fascinating property of quantum theories. 
The concept of entanglement entropy associated with a sub-region of space in quantum field theory is directly related to an observer who can only access the information of the sub-region. Tracing over the degrees of freedom outside the sub-region introduces a reduced density matrix, which can be used to compute the entanglement entropy. Exact computations of the entanglement entropy in quantum field theories are known to be difficult, even in the free field theories. Nevertheless there have been progresses in 1+1 dimensions \cite{Holzhey:1994we}\cite{Calabrese:2004eu}\cite{Calabrese:2009qy}\cite{Casini:2009sr}. 
See also \cite{Bennett:1995tk}. Sometimes, it is easier to compute the R\'enyi entropy \cite{Renyi}\cite{Klebanov:2011uf}\cite{Belin:2013uta}, that is also physically relevant. 
While the holographic approaches for computing the entropies \cite{Ryu:2006bv}\cite{Ryu:2006ef}\cite{Hubeny:2007xt}\cite{Nishioka:2009un}\cite{Casini:2011kv}\cite{Lewkowycz:2013nqa} provide valuable results, they are limited to the strongly coupled regime of the field theories. 
Thus exact and analytic results of the entanglement entropy and the R\'enyi entropy in the realistic field theory systems are valuable resources for gaining insights on the nature of quantum entanglement. 

Background gauge fields are often useful to manipulate quantum fields and to find more information about the physical systems that we are interested in. The time and space components of the 1+1 dimensional gauge potential are called chemical potential and current, respectively. The entropies of the quantum systems with finite chemical potential at zero temperature has been studied previously. The results indicate that the entropies do not depend on the chemical potential for free fermions at zero temperature  \cite{Ogawa:2011bz}\cite{Herzog:2013py} and for infinite field theory systems with a single interval \cite{CardySlides:2016}. The entanglement entropy with a current was also considered in a rather different context \cite{Arias:2014ksa}. These results are certainly interesting and deserve deeper understanding!

In a recent paper \cite{Kim:2017xoh}, we have provided the general formulas for the entanglement and R\'enyi entropies and the mutual information of 2 dimensional Dirac fermions in the presence of background gauge fields, current and chemical potential. Direct and analytic computations of them uncovered novel and interesting results in the low temperature and large radius limits. We summarize the salient results here. 

\begin{itemize}
\item First, we have shown that the entropies do depend on the chemical potential at zero temperature. This happens when the chemical potential coincides with one of the energy levels of the quantum system we consider. This is a non-trivial generalization of earlier results \cite{Ogawa:2011bz}\cite{Herzog:2013py}. This show that the entropies are useful for probing the energy spectra at zero temperature. The same properties are true for the mutual information. 

\item Second, we have shown that the entropies and the mutual (R\'enyi) information are the periodic function of the current $J$ in the low temperature limit. When we dial the modulus parameter $\alpha=2\pi \tau_1$, the current $J$ also plays the role of a `beat frequency.' We further have shown that the entropies for the periodic fermion on the spatial circle vanish at zero temperature, while those of the anti-periodic fermion have non-zero contributions. 

\item Third, in the large radius limit, the dependences of the entropies and the mutual information on the chemical potential and current vanish at least as fast as $\mathcal O \left(\ell_t/L\right)^2$, where $\ell_t$ is the size of the sub-systems that we measure the entropies inside the total system with a size $2\pi L$. This supports a recent claim that the entanglement entropy in an infinite system is independent of chemical potential $\mu$  \cite{CardySlides:2016}. We have generalized this claim for the mutual information, for the case with the multiple intervals, and for the systems with the current $J$ as well as the chemical potential $\mu$.

\item Finally, the total mutual (R\'enyi) information in general depends on the sizes of the sub-systems and their separations. Surprisingly, the current and chemical potential dependent parts of the mutual (R\'enyi) information only depend on the sub-systems sizes and are {\it independent} of the separation between the disjoint sub-systems. 

\end{itemize}

All these interesting properties are expected to present in general quantum systems and to have distinctive experimental signatures, which can be easily verifiable. Mutual (R\'enyi) information is finite and thus especially relevant for this purpose.    

The aim of this paper is to provide the details of the computations in the various limits explicitly. Here we also perform the computations in a high temperature limit, that has two non-trivial limits depending on a modulus parameter $\tau_1$ in $\tau=\tau_1 + i \tau_2$. Unlike the low temperature limit, the high temperature limit is largely determined by the combinations of parameters that are proportional to $1/\tau$. Thus in the high temperature limit, $\beta \to 0$, we have two non-trivial limits depending on $\tau_1$.   
\begin{align}
\frac{1}{\tau} = \frac{1}{\tau_1 + i \tau_2} 
\quad \rightarrow \quad \left\{ \begin{array}{lll}
-\frac{i}{ \tau_2} \;,  & \qquad  & \tau_1=0 \;, \\  
~~\frac{1}{\tau_1 }\;,  & & \tau_1 \neq 0 \;. \end{array} \right.  
\end{align} 
While the entanglement entropy for $\tau_1=0$ is usually connected to the thermal entropy, the other case can be interesting as well. Thus we compute both the cases in the high temperature limit.   
We also list the entropies and mutual information as functions of the modulus parameters without the chemical potential and current.  

The rest of the paper is organized as follows. In \S \ref{sec:PartitionFunction}, we consider the Dirac fermion in a 2 dimensional torus and present the derivation of the partition function in the presence of constant gauge fields. We did so by exploiting the equivalence between the twisted boundary conditions and the background gauge fields. Then we review a useful way to compute the entanglement entropy in \S \ref{sec:EEReview}, followed by the generalization of the entanglement entropy in the presence of chemical potential and current in \S \ref{sec:EEMuJ}. Computations for the various limits, the low temperature limit, the large radius limit, and the high temperature limit, of the entropies with chemical potential, current, both, and with only modulus parameters are presented in \S \ref{sec:EEMu}, \S \ref{sec:EEJ}, \S \ref{sec:EEMuJGeneral}, and \S \ref{sec:EEModulus}, respectively. We also compute the mutual information for those limits in \S \ref{sec:MutualInformation}. We conclude by mentioning future directions in \S \ref{sec:conclusion}.

\section{Partition function in the presence of background gauge fields} \label{sec:PartitionFunction}

We review the construction of the partition function of 1+1 dimensional Dirac fermion in the presence of the constant background gauge fields, chemical potential $\mu$ and current $J$. The basic properties can be found in \cite{DiFrancesco:1997nk}\cite{Hori:2003ic}. In particular we use the equivalence between the twisted boundary conditions and the background gauge fields to build up the partition function for the 2-dimensional Dirac fermion in the presence of the background gauge fields. 

Consider the action with the Dirac fermion $\psi$   
\begin{align}
\mathcal S = \frac{1}{2\pi} \int d^2 x ~i  \bar \psi  \gamma^\mu \left(\partial_\mu + i A_\mu \right) \psi \;, 
\end{align}
where $\mu=0, 1$ are the time and space coordinates with $\gamma^0 = \sigma_1, \gamma^1 = -i \sigma_2$ in terms of Pauli matrices and in the matrix form  
\[ \gamma^0 = \left( \begin{array}{cc}
~0~ & ~1~ \\
1 & 0 \end{array} \right)\;, \qquad 
\gamma^1 = \left( \begin{array}{cc}
~0~ & -1~ \\
1 & 0 \end{array} \right)\;,
\] 
$\bar \psi = \psi^\dagger \gamma^0$, 
and constant background gauge fields $A_0 = \mu, ~A_1 = J $ that are identified as chemical potential and current.

To compute the partition function, we consider a torus with the modular parameter $\tau = \tau_1 + i \tau_2$. Thus the space of coordinate $\zeta = \frac{1}{2\pi} (s + i t)$ is identified as $\zeta \equiv \zeta + 1 \equiv \zeta + \tau$. The spatial coordinate $s=x^1 $ is compactified with the circumference $2\pi L$ (with $L=1$ in this subsection), while $t$ is the Euclidean time with periodicity $2\pi \tau_2 = \beta = 1/T$.  
We decompose the Dirac field 
\[ \psi  = \left( \begin{array}{c}
~\psi_-~ \\
\psi_+ \end{array} \right)\;,
\] 
and consider twisted boundary conditions 
\begin{align}\label{TwistedeBC}
\psi_- (t,s) &= e^{-2\pi i a} \psi_- (t,s+2\pi) = e^{-2\pi i b} \psi_- (t+2\pi \tau_2, s+2\pi \tau_1) \;, \\
\psi_+ (t,s) &= e^{2\pi i \tilde a} \psi_+ (t,s+2\pi) = e^{2\pi i \tilde b} \psi_+ (t+2\pi \tau_2, s+2\pi \tau_1) \;. 
\end{align}
It is useful to consider an equivalent description that has the periodic Dirac fermion with the following flat gauge connection $\tilde A_\mu$ (different name because we also have the background gauge fields) on the torus   
\begin{align}\label{FlatConnection}
\tilde A = \tilde A_\mu dx^\mu = \frac{\pi i}{\tau_2} [(b-\tau a) d\bar \zeta - (\tilde b-\bar \tau \tilde a) d\zeta] \quad \to\quad  a ds + \frac{b -a\tau_1}{\tau_2} dt \;. 
\end{align}
where the arrow means to take a special case, $ \tilde a = a$ and $\tilde b = b$. Thus in this equivalent description, one can identify the relation between the background gauge fields and the twist parameters $a$ and $b$. 
\begin{align}
a = \tilde J \;,  \qquad b = \tau_1 \tilde J + i \tau_2 \tilde \mu \;.
\end{align}
These equivalences indicate that one can treat the twisted boundary condition and the constant background gauge fields on an equal footing. Once the partition function with the twisted parameters are constructed, one can readily generalize that in the presence of the constant background gauge fields. 

The partition function is a trace of the states of the Hilbert space that are constructed using the twisted boundary condition on $s \sim s+2\pi$ along with the Euclidean time evolution $t \to t+2\pi \tau_2$ represented by the operator $ e^{-2\pi \tau_2 H}$, where $H$ is Hamiltonian. The latter also induces the space translation $s \to s - 2\pi \tau_1$ represented by the operator $ e^{-2\pi i \tau_1 P}$ (with momentum operator $P $) together with phase rotation due to the presence of fermion $ e^{-2\pi i (b-1/2) F_A}$, where $F_A = \frac{1}{2\pi} \int ds ( \psi_+^\dagger \psi_+ - \psi_-^\dagger \psi_-)$ is the fermion number. Thus 
\begin{align}
Z_{[a,b]} = Tr \left(e^{-2\pi i (b-1/2) F_A} e^{-2\pi i \tau_1 P} e^{-2\pi \tau_2 H} \right)  \;.
\end{align}   
The last two factors can be rewritten into a useful form $e^{-2\pi i \tau_1 P} e^{-2\pi \tau_2 H} 
= q^{(H-P)/2} \bar q^{(H+P)/2} $ with $ q= e^{2\pi i \tau}$. The partition function can be written 
\begin{align}
Z_{[a,b]} = \left|\eta(\tau) \right|^{-2} \left|\vartheta \left[\substack{-a+1/2 \\ b-1/2 } \right] (0|\tau)\right|^2
 \;,
\end{align} 
in terms of the Jacobi theta functions $\vartheta \left[\substack{\alpha \\ \beta} \right]$ and the Dedekind $\eta(\tau)$ function  
\begin{align}\label{ThetaFunctionSeriesForm}
\vartheta \left[\substack{\alpha \\ \beta} \right] (z|\tau) = \sum_{n \in \mathbb Z} q^{(n + \alpha)^2/2} e^{2\pi i (z+\beta)(n + \alpha)} \;, \qquad 
\eta(\tau) = q^{1/24} \prod_{n=1}^\infty (1 - q^n) \;.
\end{align} 
Focusing on the NS-NS sector, $a=1/2, b=1/2$, with the anti-periodic boundary conditions of spatial and time circles, the partition function has the form 
\begin{align}
Z_{[1/2,1/2]} = \left|\eta(\tau) \right|^{-2} \Big|\sum_{n \in \mathbb Z} q^{n^2/2} \Big|^2
 \;. 
\end{align}
In the literature, the following notations are also used $\vartheta_3(z|\tau) = \vartheta \left[\substack{0 \\ 0} \right] (z|\tau), \vartheta_2(z|\tau) = \vartheta \left[\substack{1/2 \\ 0} \right] (z|\tau)$, $\vartheta_4(z|\tau) = \vartheta \left[\substack{0 \\ 1/2} \right] (z|\tau), \vartheta_1(z|\tau) = \vartheta \left[\substack{1/2 \\ 1/2} \right] (z|\tau) $. $\vartheta_2(z|\tau)$ is related to the periodic spatial circle and anti-periodic time circle. Thus we focus on $\vartheta_3$ and $\vartheta_2$.

In the presence of the chemical potential $\mu$ and the current $J$, the partition function has the general form 
\begin{align}
Z_{[a,b]} (\mu, J) = Tr \left(e^{2\pi i (\tau_1 J + i \tau_2 \mu + b-1/2) F_A} e^{-2\pi i \tau_1 P} e^{-2\pi \tau_2 H} \right)  \;.
\end{align}   
This can be understood by examining the Dirac action in the presence of the background gauge fields as well as the effective flat gauge connection $\tilde A$ given in \eqref{FlatConnection}. 
\begin{align}
\tilde {\mathcal S} = \frac{1}{2\pi} \int d^2 x ~i  \bar \psi  \gamma^\mu \left(\partial_\mu + i A_\mu + i \tilde A_\mu\right) \psi \;, 
\end{align}
where the Dirac field has the periodic boundary condition.  
In particular, the mode expansion of the fermion field depends not only on the twisted boundary condition, but also on the presence of the current $J$. For example, the fields obeying the twisted boundary condition \eqref{TwistedeBC} is further modified in the presence of the current $J$:
\begin{align}
\psi_- = \sum_{r \in \mathbb Z + a} \psi_r (t) e^{irs} \quad \to \quad \sum_{\tilde r \in \mathbb Z + a + J} \psi_r (t) e^{i \tilde r s} \;.
\end{align}
Thus the presence of current changes the periodicity of the compact fermions, and thus periodic fermion is no longer periodic in the presence of current.  
In terms of the Jacobi theta functions, 
\begin{align}
Z_{[a,b]} (\mu, J) 
&=\left|\eta(\tau) \right|^{-2} \left|\vartheta \left[\substack{-a_{\! J} +1/2 \\ b_{\mu,\! J}-1/2 } \right] (0|\tau)\right|^2
=\left|\eta(\tau) \right|^{-2} \left|\vartheta \left[\substack{-a_{\! J}+1/2 \\ b_{\! J}-1/2 } \right] (i\tau_2 \mu|\tau)\right|^2 \nonumber \\
&=\left|\eta(\tau) \right|^{-2} \left|\vartheta \left[\substack{-a_{\! J}+1/2 \\ b-1/2 } \right] (\tau_1 J +i\tau_2 \mu|\tau)\right|^2
 \;,
\end{align} 
where 
\begin{align}
a_{\! J} = a + J \;, \qquad b_{\mu,\!J} = b_{\! J} + i \tau_2 \mu = b + \tau_1 J + i \tau_2 \mu \;.
\end{align}
The periodicity of temporal direction is not modified. This is consistent with imposing the anti-periodic boundary condition on the temporal direction.   

\section{Entropies in the presence of background gauge fields}\label{sec:EE}

In this chapter, we first review the construction of the entanglement and R\'enyi entropies using the Replica trick \cite{Calabrese:2004eu}\cite{Calabrese:2009qy}\cite{Casini:2009sr}, followed by their generalization in the presence of the twisted boundary conditions and also background gauge fields. We then perform detailed computations of the entropies in the various limits. 

\subsection{Basics} \label{sec:EEReview}

Let us consider a reduced density matrix for a free Dirac fermion $\psi$ on an Euclidean plane by tracing its vacuum state over the degrees of freedom lying outside a given set of disjoint intervals $ (u_{\tilde a},v_{\tilde a})$, $\tilde a=1,...,p$. The density matrix $\rho (\psi_{in},\psi_{out})$ can be written as a functional integral on an Euclidean plane with appropriate boundary conditions $\psi =\psi_{in}$, and $\psi =\psi_{out}$ along each side of the cuts $(u_{\tilde a},v_{\tilde a})$ 
\begin{align}
\rho (\psi_{in},\psi_{out}) = \frac{1}{Z[1]}\int D\psi e^{-S[\psi]} \;,
\end{align}
where $Z[1]$ is a normalization factor and gives the density matrix $Tr (\rho) =1$. To evaluate the R\'enyi entropy, we consider $Tr(\rho^{n})$ with $n$ copies of the cut plane by sewing together a cut $(u_{\tilde a},v_{\tilde a})_o^{k}$ with the next cut $(u_{\tilde a},v_{\tilde a})_i^{k+1}$, for all cuts $\tilde a= 1 ,..., p$, and the replica copies $ k=1, ..., n$. The copy $n+1$ coincides with the first one. The trace of $\rho^{n}$ is $Z[n]$ for the fields, and thus 
\begin{align}
Tr(\rho^{n}) = Z[n]/Z[1]^n \;. 
\end{align}

Following \cite{Casini:2005rm}, we consider $n$-copies of fermions on a single Riemann surface instead of a fermion on $n$-copies of Riemann surfaces. 
\begin{align}
\psi=\left(\begin{array}{c} 
\psi_{1}(x) \\ 
\vdots \\  
\psi_{n}(x) \end{array} \right) \;,
\end{align}
where $\psi_k (x)$ represents the field on the $k$-th copy. For a fermion, we need to introduce a $-$ sign for a trace, which is required when we connect the last copy to the first one. Furthermore, each fermion copy requires another $-$ sign due to a $2\pi$ rotation before connecting to the next copy \cite{Kabat:1995eq}. Putting them together for a single cut $(u_{\tilde a},v_{\tilde a})$ with the $n$-replica copies, from the first copy $(u_{\tilde a},v_{\tilde a})_{in}^1$ to the last one $(u_{\tilde a},v_{\tilde a})_{out}^{n+1}$, requires $(-1)^{n+1}$ factor. In this way, performing the trace for each cut can be described by multiplying the matrix  
\[
T = \left( \begin{array}{ccccc}
0 & ~1~ & ~ ~ & ~ ~  &  ~ ~ \\
& 0 & 1 &  &  \\
&  & . & . &  \\
&  &  & 0 & 1 \\
(-1)^{(n+1)} &  &  &  & 0
\end{array}\right) \]
for $u_{\tilde a}$ located on one end of a cut and $T^{-1}$ for $v_{\tilde a}$ located on the other end of the cut. By changing a basis by a unitary transformation in the replica space, one can diagonalize $T$. The corresponding eigenvalues of the $T$ are $e^{i\frac{k}{n}2\pi }$, where $k=-\frac{(n-1)}{2}$, $- \frac{(n-1)}{2}+1$, ..., $\frac{(n-1)}{2}$. This reveals that the space is simply connected but the field $\psi$ is not singled valued. By diagonalizing the $T$ matrix, the problem is reduced to $n$ decoupled fields $\psi^{k}$ living on a single plane. These fields are multivalued, since when encircling $u_{\tilde a}$ or $v_{\tilde a}$ they are multiplied by $e^{i\frac{k}{n}2\pi}$ or $e^{-i\frac{k}{n}2\pi}$, respectively.

The multivaluedness can be removed by introducing an external constant gauge field coupled with the fields  $\psi^{k}$, which is single-valued. This can be described by the Lagrangian density
\begin{align}
\mathcal L_{k} = i \bar{\psi}^{k} \gamma ^{\mu } \left( \partial_{\mu } + i \tilde A_{\mu}^{k} \right) \psi^{k} 
\;.
\end{align}
The singular gauge transformation, $\psi^{k}(x)\to \exp \left( -i \int_{x_{0}}^{x} dx^{'\mu } \tilde A_{\mu }^{k} (x^{'})\right) \psi^{k}\left( x\right)$, would get rid of the gauge field $\tilde A_{\mu}^{k}$, leaving the fields to be multivalued. Thus the proper boundary conditions for the $\tilde a$-th cut on the fields $\psi^{k}$ are 
\begin{align}
\oint_{C_{u_{\tilde a}}}dx^{\mu } \tilde A_{\mu }^{k}(x) &= -\frac{2 \pi k}{n} \;,  \qquad\qquad  
\oint_{C_{v_{\tilde a}}}dx^{\mu } \tilde A_{\mu }^{k}(x) = \frac{2 \pi k}{n} \;.
\end{align}
This holds for any two circles going around $u_{\tilde a}$ and $v_{\tilde a}$. Putting together, we get 
\begin{align}
\epsilon^{\mu\nu}\partial_{\nu} \tilde A_{\mu}^{k}(x) = 2\pi \frac{k}{n}
\sum_{\tilde a=1}^{p} \left[\delta (x-u_{\tilde a})-\delta (x-v_{\tilde a}) \right] \;. 
\end{align}

The R\'enyi entropy can be evaluated with $Z[n] = \prod_{k=-\frac{(n-1)}{2}}^{\frac{n-1}{2}} Z_k$, where $Z_{k}$ can be obtained by the vacuum expectation value in the free Dirac theory 
\begin{align}
Z_{k} = \langle e^{i\int \tilde A_{\mu}^{k} J_{k}^{\mu}d^{2}x} \rangle \;,
\end{align}
where $J_{k}^{\mu }$ and $\tilde A_{\mu}^{k}$ are is the Dirac current and background gauge field. 
$Z_{k}$ can be evaluated using the bosonization technique. Then the current is 
\begin{align}
J_{k}^{\mu }\to \frac{1}{2\pi}\epsilon ^{\mu \nu }\partial_{\nu}\phi \;,  
\end{align}
where $\phi $ is a real scalar field. For a free massless Dirac field, 
$\mathcal L_{\phi}=\frac{1}{2}\partial _{\mu }\phi \partial ^{\mu }\phi$. Thus 
\begin{align}
Z_{k}= \langle e^{i\int \tilde A_{\mu }^{k}\frac{1}{2\pi}\epsilon ^{\mu
\nu }\partial _{\nu}\phi d^{2}x}\rangle 
=\left\langle e^{-i \frac{k}{n} \sum_{\tilde a=1}^{p} \left(\phi (u_{\tilde a})-\phi (v_{\tilde a})\right)
}\right\rangle
=\prod_{\tilde a=1}^p \langle \sigma_k (u_{\tilde a}) \sigma_{-k} (v_{\tilde a}) \rangle \;,
\end{align}
where the vacuum expectation values correspond to those of the massless Dirac theory, and $\sigma_k (u_{\tilde a})$ and $\sigma_{-k} (v_{\tilde a})$ are the twist operators with conformal dimension $ \frac{k^2}{2 n^2}$ in the $Z_n$ orbifold theory for free Dirac fermion \cite{Azeyanagi:2007bj}. 

Since ${\cal L}_\phi$ is quadratic in $\phi$, one can perform the path integral for the field $\phi$. 
\begin{align}
\langle e^{-i\int f(x)\phi (x)d^{2}x} \rangle = e^{-\frac{1}{2}\int f(x)G(x-y)f(y)d^{2}xd^{2}y} \;,
\end{align}
with the correlator $G(x-y)=-\frac{1}{2\pi}\log \left|x-y\right| $. Thus
\begin{align}\label{Ren1}
\log Z_{k} &= \frac{2k^2}{n^2} \log \left|\frac{\prod_{\tilde a<\tilde b} (u_{\tilde a}-u_{\tilde b})(v_{\tilde a}-v_{\tilde b})}{\prod_{\tilde a,\tilde b} (u_{\tilde a}-v_{\tilde b})} \varepsilon ^p \right|  \;,
\end{align}
where $\varepsilon $ is a cutoff introduced to split the coincidence points, $\left| u_{\tilde a}-u_{\tilde a}\right|$, $\left| v_{\tilde a}-v_{\tilde a}\right| \to \varepsilon$. Thus the resulting entropies diverge as $\varepsilon \to 0$. Summing over $k$, we obtain
\begin{align}
S_n &= \frac{1}{1-n}\log (Tr(\rho ^{n})) =\frac{1}{1-n}\sum_{k=-\frac{n-1}{2}}^{\frac{n-1}{2}}\log Z_{k} \;, \\
S &= -\frac{1}{3} \log \left|\frac{\prod_{\tilde a<\tilde b} (u_{\tilde a}-u_{\tilde b})(v_{\tilde a}-v_{\tilde b})}{\prod_{\tilde a,\tilde b} (u_{\tilde a}-v_{\tilde b})} \varepsilon ^p \right| \;. 
\label{ent1a}
\end{align}
This agrees exactly with the general formula for the entanglement entropy for conformal theories obtained in \cite{Calabrese:2004eu}.

This basic review teaches us to consider the $Z_n$ orbifold theory for the free Dirac fermion and the correlation functions of the corresponding twist operators \cite{Azeyanagi:2007bj} to compute the entropies. We are going to generalize this result to the case with the background gauge fields on the 2 dimensional torus. For the rest of the paper, we focus on a single interval. The generalization for the multiple intervals is straightforward \cite{Herzog:2013py}.

\subsection{Generalizations with background gauge fields} \label{sec:EEMuJ}

We consider two point functions of the $k-$th twist operators $\sigma_k (u)$ and $\sigma_{-k} (v)$ with conformal dimension $ \frac{k^2}{2 n^2}$ in the $Z_n$ orbifold theory for free Dirac fermion on 2-dimensional torus with twisted boundary conditions parametrized by $a$ and $b$ introduced in \eqref{TwistedeBC} \cite{DiFrancesco:1997nk}\cite{Azeyanagi:2007bj}\cite{Herzog:2013py}. 
\begin{align}
\langle \sigma_k (u) \sigma_{-k} (v) \rangle_{a,b} = \Big| \frac{2\pi \eta (\tau)^3 }{\vartheta [\substack{1/2 \\ 1/2 }](\frac{u-v}{2\pi L}|\tau)} \Big|^{2\frac{k^2}{n^2}} ~
\Big| \frac{\vartheta [\substack{1/2-a \\ b-1/2 }](\frac{k}{n} \frac{u-v}{2\pi L}|\tau)}{\vartheta [\substack{1/2-a \\ b-1/2 }](0|\tau)} \Big|^2 \;, 
\end{align} 
where the first term is the generalization of the results in the previous section, while the second one comes into play because of the torus and the corresponding spin structures.

Following \S \ref{sec:PartitionFunction}, we generalize the two point functions of the twist operators in the presence of the current $J$ and the chemical potential $\mu$. 
\begin{align}\label{CorrelatorTwistedOp}
\langle \sigma_k (u) \sigma_{-k} (v) \rangle_{a,b,J,\mu} = \Big| \frac{2\pi \eta (\tau)^3 }{\vartheta [\substack{1/2 \\ 1/2 }](\frac{u-v}{2\pi L}|\tau)} \Big|^{2\frac{k^2}{n^2}}~ 
\Big| \frac{\vartheta [\substack{1/2-a-J \\ b-1/2 }](\frac{k}{n} \frac{u-v}{2\pi L} + \tau_1 J + i \tau_2 \mu|\tau)}{\vartheta [\substack{1/2-a-J \\ b-1/2 }](\tau_1 J + i \tau_2 \mu|\tau)} \Big|^2 \;. 
\end{align}
The first factor is independent of the twisted boundary conditions and the background gauge fields. 

Let us consider a subsystem $A$, whose size is given by $(u - v)/2\pi L $. Using the replica trick, we get the expression for the R\'enyi entropy  
\begin{align}\label{GeneralEEFormula}
S_n = \frac{1}{1- n} \left[ \log \text{Tr}(\rho_A)^n \right] 
= \frac{1}{1- n} \bigg[ \log \prod_{k=-\frac{n-1}{2}}^{\frac{n-1}{2}} \langle \sigma_k (u) \sigma_{-k} (v) \rangle_{a,b,J,\mu} \bigg]  = S_n^0 + S_n^{\mu,J} \;.
\end{align}
The entropies factorizes into two parts, $S_n^0$ and $S_n^{\mu,J}$. $S_n^0$ is independent of the spin structures or background gauge fields. 
\begin{align}
S_n^0 = \frac{1}{1- n} \bigg[  \sum_{k=-\frac{n-1}{2}}^{\frac{n-1}{2}} \frac{k^2}{n^2} \log \Big|\frac{2\pi \eta (\tau)^3 }{\vartheta [\substack{1/2 \\ 1/2 }](\frac{u-v}{2\pi L}|\tau)} \Big|^2 \bigg] = - \frac{n+1}{12n} \log \bigg|\frac{2\pi \eta (\tau)^3 }{\vartheta [\substack{1/2 \\ 1/2 }](\frac{u-v}{2\pi L}|\tau)} \bigg|^2 \;.
\end{align}
We use the sum $ \sum_{k=-\frac{n-1}{2}}^{\frac{n-1}{2}} \frac{k^2}{n^2} = \frac{n^2 -1}{12 n}$ that can be done straightforwardly. This contribution $S_n^0$ has been analyzed before. 

Below we focus on $S_n^{\mu,J}$ that depends on the chemical potential and current 
\begin{align}\label{GeneralEEFormulaJmu}
S_n^{\mu,J}
= \frac{1}{1- n} \bigg[ \sum_{k=-\frac{n-1}{2}}^{\frac{n-1}{2}} \log \Big| \frac{\vartheta [\substack{1/2-a-J \\ b-1/2 }](\frac{k}{n} \frac{u-v}{2\pi L} + \tau_1 J + i \tau_2 \mu|\tau)}{\vartheta [\substack{1/2-a-J \\ b-1/2 }](\tau_1 J + i \tau_2 \mu|\tau)} \Big|^2 \bigg]   \;.
\end{align} 
We note that the Jacobi theta function has $k$ dependence in the argument, and thus summing over $k$ and taking $n\to 1$ limit to get the entanglement entropy are non-trivial. We perform explicit computations in various limits, such as the zero temperature limit and the large radius limit, to show novel and interesting results. 

For the high temperature limit, one needs a different form that can be obtained using a modular transform. One can show the following identity using the Poisson resummation 
\begin{align}
\vartheta \left[\substack{\alpha \\ \beta} \right] ({z}/{\tau}|-{1}/{\tau}) = (-i \tau)^{1/2} e^{\pi i z^2/\tau + 2\pi i \alpha \beta} \vartheta \left[\substack{\beta \\ -\alpha} \right] (z|\tau) \;.
\end{align} 
Using this one can recast the entropy formula \eqref{GeneralEEFormulaJmu} as 
\begin{align}\label{GeneralEEFormulaJmuHighT}
\tilde S_n^{\mu,J}
= \frac{1}{1 \- n} \sum_{k=-\frac{n-1}{2}}^{\frac{n-1}{2}} \log \Big|\frac{e^{-i\frac{\pi}{\tau} (\frac{k}{n}\frac{u-v}{2\pi L}+ \tau_1 J + i \tau_2 \mu)^2}}{e^{-i\frac{\pi}{\tau} ( \tau_1 J + i \tau_2 \mu)^2}} \frac{\vartheta [\substack{1/2-b \\ 1/2-a-J }](\frac{k}{n} \frac{u-v}{2\pi L} \frac{1}{\tau} + \frac{\tau_1 J + i \tau_2 \mu}{\tau}|-\frac{1}{\tau})}{\vartheta [\substack{1/2-b \\ 1/2-a-J }](\frac{\tau_1 J + i \tau_2 \mu}{\tau}|-\frac{1}{\tau})} \Big|^2  \;.
\end{align} 
Note that $1/\tau$ is everywhere that plays crucial role in the high temperature limit. In particular, there are two non-trivial limits depending on the presence of $\tau_1$. The details can be found below. 

Now we have developed the general formula for the Entanglement and R\'enyi entropies. We are going to perform the computations in the presence of chemical potential in \S \ref{sec:EEMu} and provide the salient features in that limit. Then we focus on the role of the current $J$ for the entropies in \S \ref{sec:EEJ}, followed by the discussion with both $\mu$ and $J$ in \S \ref{sec:EEMuJGeneral}. To be complete, we also provide the results for $\mu =J=0 $ in \S \ref{sec:EEModulus}. The computations in \S \ref{sec:EEMu} are thorough, while those in the other sections are in general brief or omitted because they are similar. 

Before moving on, let us clarify some of the notations. There are two different periodicity for both the spatial and temporal directions. We call the NS sector for the fermion with the anti-periodic boundary condition and the R sector for the case with the periodic boundary condition.  In this paper we only consider the anti-periodic boundary condition for the temporal direction, NS sector, that is to assign $b=1/2$. 
The NSNS sector means to have the anti-periodicity for both directions, while the RNS sector to be periodic for the spatial direction and anti-periodic for the temporal direction.

\subsection{Chemical potential}\label{sec:EEMu}

In this section, we consider a more familiar case with the chemical potential $\mu$ by setting $J=0$ and one of the modulus parameter $\alpha=2\pi \tau_1 = 0$. We present the details in the low temperature limit, the large radius limit, and the high temperature limit in turn. We only consider the anti-periodic condition of the Dirac fermion for the temporal circle.

\subsubsection{Low temperature limit}

It has been shown that entropies at zero temperature are independent of a finite chemical potential \cite{Ogawa:2011bz}\cite{Herzog:2013py}. It turns out that there are more on the story. Here we show that there are more interesting and refined zero temperature limits, $\beta \to \infty$, $\mu \to N/2$ keeping $\beta (\mu - N/2) \to const.$. Then the entropies actually depend on the chemical potential $\mu$ at zero temperature. We identify $N/2$ as the energy levels of the Dirac fermion on a circle. Thus, when the chemical potential hits the energy eigenvalues of a theory at zero temperature, the entropies depending on the chemical potential do not vanish. This section contains the details.

The R\'enyi entropy $S_n^{\mu}$, \eqref{GeneralEEFormulaJmu}, reduces to the known form for the anti-periodic boundary condition for a spatial circle, NS-sector, with $a=1/2$ \cite{Ogawa:2011bz}\cite{Herzog:2013py}. 
\begin{align}\label{SAmu3}
S_n^\mu &=  \frac{1}{1- n}\sum_{k=-\frac{n-1}{2}}^{\frac{n-1}{2}} \log \Big|\frac{\vartheta_3 (\frac{k}{n} \frac{\ell_t}{2\pi L}+\frac{i\beta\mu}{2\pi} |i\beta )}{\vartheta_3 (\frac{i\beta\mu}{2\pi} |i\beta)} \Big|^2  \;, 
\end{align} 
where we use the notation $\beta = 2\pi \tau_2$ and $\ell_t = u-v $. The case with multiple intervals is straightforward to generalize by replacing the sub-system sizes as $\ell_t = \sum_{a=1}^p (u_a -v_a) $. (See {\it e.g.} \cite{Herzog:2013py}). 

Using the product representation 
\begin{align}\label{Theta3}
\vartheta_3 (z|\tau) &= \prod_{m=1}^\infty (1 - q^m)(1 + y q^{m-1/2})(1 + y^{-1} q^{m-1/2}) \;, \qquad 
q=e^{2\pi i \tau} ~~\&~~ y = e^{2\pi i z} \;,
\end{align} 
along with $y_1 = e^{- \beta\mu + 2\pi i \frac{k}{n} \frac{\ell_t}{2\pi L}}, ~y_2 = e^{- \beta\mu}, ~ q=e^{-\beta}$, we compute the R\'enyi entropy in the low temperature limit, $\beta \to \infty$, 
\begin{align}\label{RenyiSAmuNS}
S_n^\mu &= \frac{1}{1\- n}  \sum_{k=-\frac{n-1}{2}}^{\frac{n-1}{2}} \log \Big|\prod_{m=1}^{\infty} \frac{(1 \- q^m)(1 \+ y_1 q^{m-1/2})(1 \+ y_1^{-1} q^{m-1/2})}{(1 \- q^m)(1 \+ y_2 q^{m-1/2})(1 \+ y_2^{-1} q^{m-1/2})} \Big|^2  \nonumber \\
&=\frac{1}{1\- n} \! \sum_{k=-\frac{n-1}{2}}^{\frac{n-1}{2}}  \sum_{m,l=1}^{\infty} \!\!  \frac{(-1)^{l-1}}{l} \left[ [y_1^l \+ y_1^{l*} \- y_2^l \- y_2^{l*}] q^{l(m-\frac{1}{2})} \+ [y_1^{-l} + y_1^{-l*} \- y_2^{-l}  \- y_2^{-l*} ]q^{l(m-\frac{1}{2})}  \right]  \nonumber \\
&= \frac{2}{1\- n}  \sum_{k=-\frac{n-1}{2}}^{\frac{n-1}{2}} \sum_{l,m=1}^{\infty} \!\! \frac{(-1)^{l-1}}{l} 
\left( e^{-l\beta\mu} + e^{l\beta\mu}\right) e^{- l\beta(m-1/2)} \left[ \cos \left( \frac{k}{n} \frac{\ell_t}{L}l\right) -1 \right]  \nonumber  \\
&=2\sum_{l=1}^{\infty} \frac{(-1)^{l-1}}{l} 
\frac{\cosh (l\beta\mu)}{\sinh ( l\beta/2 )} \frac{1}{1\- n} \left[   \sin \left( \frac{\ell_t}{2L}l\right) \csc \left(\frac{1}{n} \frac{\ell_t}{2L}l\right) \-n\right] \;. 
\end{align} 
In the second line, we use the Taylor expansion for the log function using the conditions $y_{1,2} q^{m-1/2}<1$ and $y_{1,2}^{-1} q^{m-1/2} < 1 $. This is an exact result, while it is an infinite sum.  

One can obtain the entanglement entropy by taking the limit $n\to 1$. 
\begin{align}\label{SAmuNS}
S^\mu &=2\sum_{l=1}^{\infty} \frac{(-1)^{l-1}}{l} 
\frac{\cosh (l\beta\mu)}{\sinh ( l\beta/2 )} \left[1- l \frac{\ell_t}{2L} \cot \left(l \frac{\ell_t}{2L} \right) \right]  \;. 
\end{align} 
Note that we expand the log terms with the (strongest) conditions $|y_{1,2} q^{1/2}| < 1$ and $|y_{1,2}^{-1} q^{1/2}| < 1$ to get the result. They translate into 
\begin{align}
e^{-\beta\mu-\beta/2} < 1   \qquad \& \qquad e^{\beta\mu-\beta/2} < 1  
\end{align}
and are satisfied for $-1/2< \mu <1/2$. Thus we see that both the entropies, the entanglement entropy and the R\'enyi entropy, vanish at the zero temperature limit. Thus, the entropies seem to be the same as those without the chemical potential. Nevertheless, this conclusion is not valid for $\mu = \pm 1/2$. We carefully look into the special values. 

For this purpose, we consider $\beta(\mu-1/2)=M=const.$ in the limit $\beta \to \infty$ and $\mu \to 1/2$.  
The product representation for $\vartheta_3$ can have a modified expansion 
\begin{align}
\vartheta_3 (z|\tau) 
&= (1 + y^{-1} q^{1/2}) \prod_{m=1}^\infty (1 - q^m)(1 + y q^{m-1/2})(1 + y^{-1} q^{m+1/2})\;. 
\end{align} 
The front factor is finite. 
Thus the R\'enyi entropy is modified to 
\begin{align}\label{RenyiFinite1}
S_n^\mu &= \frac{1}{1\- n} \sum_{k=-\frac{n-1}{2}}^{\frac{n-1}{2}} \log \Big|\frac{(1 + y_1^{-1} q^{1/2})}{(1 + y_2^{-1} q^{1/2})} \prod_{m=1}^{\infty} \frac{(1 - q^m)(1 + y_1 q^{m-1/2})(1 + y_1^{-1} q^{m+1/2})}{(1 - q^m)(1 + y_2 q^{m-1/2})(1 + y_2^{-1} q^{m+1/2})} \Big|^2 
\;. 
\end{align} 
The first part inside the log can be evaluated without assumptions in terms of Pochhammer symbols, that is written in the appendix \S \ref{sec:AppendixFPEAZT}. For simplicity, we assume that $e^M<1$ to evaluate the finite part with $y_2^{-1} q^{1/2} = e^{\beta (\mu-1/2)}= e^M$ in the Taylor series. If one wants to consider the other case, $e^M>1$, that can be done similarly. The second part in the log can be evaluated in a straightforward manner. To do so, note that the following condition $e^{-\beta\mu- \beta/2} < 1$ and $e^{\beta\mu-3 \beta/2} < 1$ can be used to have an expansion at the zero temperature limit. Those conditions are satisfied for 
$$
-1/2< \mu <3/2 \;.
$$ 
Thus one can set $\mu \to 1/2$ in the entropy formula, and that is more appropriate. Thus we get  
\begin{align}\label{EEFinteAtZeroT}
S_n^\mu  &=2\sum_{l=1}^{\infty} \frac{(-1)^{l-1}}{l} 
\left[e^{lM}+ \frac{e^{ -l\beta /2}}{\sinh (l\beta/2)}  \right] \frac{1}{1\- n} \left[   \sin \left(\frac{\ell_t}{2L}l\right) \csc \left(\frac{1}{n} \frac{\ell_t}{2L}l\right) \-n\right] 
\;. 
\end{align} 
The corresponding entanglement entropy is  
\begin{align}\label{RenyiEFinteAtZeroT}
S^\mu &=2\sum_{l=1}^{\infty} \frac{(-1)^{l-1}}{l} 
\left[e^{lM}+ \frac{e^{ -l\beta /2}}{\sinh (l\beta/2)} \right] \left[1- l \frac{\ell_t}{2L} \cot \left(l \frac{\ell_t}{2L} \right) \right]  
\;. 
\end{align} 

This is very interesting! The results of the entanglement entropy \eqref{EEFinteAtZeroT} and R\'enyi entropy \eqref{RenyiEFinteAtZeroT} reveal that there is a finite contribution for $\beta(\mu-1/2) =M= const.$ as $\beta \to \infty$ and $\mu \to 1/2$. 
\begin{align}\label{Nu3ZeroTHalfMu}
S_{\text{finite}}^\mu &= 2\sum_{l=1}^{\infty} \frac{(-1)^{l-1}}{l}  e^{lM} \left[1- l \frac{\ell_t}{2L} \cot \left(l \frac{\ell_t}{2L} \right) \right]  
\;. 
\end{align} 
We identify the chemical potential $\mu (=1/2)$ as one of the energy levels of the Dirac fermion on a circle. We repeat the computations to see the similar results for other values of the chemical potential that coincide with the energy levels of the Dirac fermion as  
\begin{align}
\beta \left(\mu - \frac{2N+1}{2} \right) = const. \;, \qquad \beta \to \infty ~~\&~~ \mu \to \frac{2N+1}{2} \;.
\end{align}

Let us turn our attention to consider the periodic boundary condition on the spatial circle, R-sector, with $a=0$. The R\'enyi entropy has the form 
\begin{align}\label{EEWithTheta2}
S_n^\mu &= \frac{1}{1- n} \sum_{k=-\frac{n-1}{2}}^{\frac{n-1}{2}} \log \Big|\frac{\vartheta_2 (\frac{k}{n} \frac{\ell_t}{2\pi L}+\frac{i\beta\mu}{2\pi} |i\beta )}{\vartheta_2 (\frac{i\beta\mu}{2\pi} |i\beta)} \Big|^2 
\end{align} 
with the theta function representation 
\begin{align}\label{Theta2}
\vartheta_2 (z|\tau) &= 2 e^{\pi i \tau/4} \cos(\pi z) \prod_{m=1}^\infty (1 - q^m)(1 + y q^m)(1 + y^{-1}q^m) \;. 
\end{align} 
Similar computations show for the entanglement entropy  
\begin{align}\label{SAmuR}
S^\mu &= \frac{1}{1- n} \bigg[ \sum_{k=-\frac{n-1}{2}}^{\frac{n-1}{2}} \log \Big|\frac{e^{-\beta \pi/4} (y_1^{1/2} + y_1^{-1/2})}{e^{-\beta \pi/4} (y_2^{1/2} + y_2^{-1/2})} \prod_{m=1}^{\infty} \frac{(1 - q^m)(1 + y_1 q^{m})(1 + y_1^{-1} q^{m})}{(1 - q^m)(1 + y_2 q^{m})(1 + y_2^{-1} q^{m})} \Big|^2 \bigg] \bigg|_{n=1} \nonumber \\
&=2\sum_{l=1}^{\infty} \frac{(-1)^{l-1}}{l} 
\left(\frac{\cosh\left(l\beta\mu\right)}{\sinh\left(l\beta/2\right)} e^{-l \beta/2} + e^{-l\beta\mu} \right) \left[1- l \frac{\ell_t}{2L} \cot \left(l \frac{\ell_t}{2L} \right) \right]  
\;. 
\end{align} 
The computation is valid for $0 \leq \mu < 1 $. We check that the second contribution in the round bracket is non-zero for $\beta \mu = const.$ in the limit $\beta \to \infty$ and $\mu \to 0$. In fact there are non-zero contributions in the entanglement entropy 
when 
\begin{align}
\beta(\mu-N) = const. \;, \qquad \beta \to \infty ~~\&~~ \mu \to N \;.
\end{align}
The integer $N$ is identified as one of the energy levels of the Dirac fermion of a compact circle with the periodic boundary condition.  

Combining the NS and R sectors together, we find the entropies acquire non-zero contribution when 
\begin{align}
\beta \left(\mu-\frac{N}{2} \right) = const. \;, \qquad \beta \to \infty ~~\&~~ \mu \to \frac{N}{2} \;,
\end{align}
for an integer $N$. 
Thus we show that the entanglement entropy at the zero temperature limit has the ability to detect the energy levels of the underlying theory. We expect this would happen generically, providing a useful way to probe the energy levels with a varying chemical potential. For example, consider the anti-periodic fermion in the R-sector at zero temperature. As one increases the chemical potential, the entanglement entropy picks up a non-zero value each time the chemical potential passes through the energy level of the system. It will be interesting to verify these features experimentally.

\subsubsection{Large radius limit} 

It has been argued that the entanglement entropy is independent of chemical potential for the infinitely long space \cite{CardySlides:2016} based on symmetry argument in 2 dimensional field theory. In this section we support and generalize the claim by evaluating the entanglement entropy by taking the limit $\frac{\ell_t}{L} \to 0 $, which can be considered as either an infinite space limit or a limit of small systems size. 

We first consider the fermion in the R-sector with a periodic boundary condition. The entropy formula  \eqref{GeneralEEFormulaJmu} reduces to \eqref{EEWithTheta2}
\begin{align}
S_n^\mu &= \frac{1}{1- n} \sum_{k=-\frac{n-1}{2}}^{\frac{n-1}{2}} \log \Big|\frac{\vartheta_2 (\frac{k}{n} \frac{\ell_t}{2\pi L}+\frac{i\beta\mu}{2\pi} |i\frac{\beta}{2\pi})}{\vartheta_2 (\frac{i\beta\mu}{2\pi} |i\frac{\beta}{2\pi})} \Big|^2 \;.
\end{align} 
A slightly modified representation for the theta function \eqref{Theta2} 
\begin{align}
\vartheta_2 (z|\tau) & =  2 e^{\pi i \tau/4} \cos(\pi z) \prod_{m=1}^\infty (1 - q^m)(1 + q^{2m} + 2 \cos (2\pi z) q^m) 
\end{align} 
is useful along with the identifications $z_1 = i\frac{\beta\mu}{2\pi} + \frac{k}{n} \frac{\ell_t}{2\pi L}, ~z_2 = i\frac{\beta\mu}{2\pi}, ~ q=e^{-\beta}$.
One can have the following expansion for $\frac{\ell_t}{L} \ll 1 $
\begin{align}
&1 + q^{2m} + 2 \cos (2\pi z_1) q^m \nonumber \\
&=  
1 + q^{2m} + 2 \cos (2\pi z_2) q^m -2 q^m \left(i \frac{k}{n} \frac{\ell_t}{L} \sinh ( \beta\mu)  +\frac{1}{2} (\frac{k}{n} \frac{\ell_t}{L})^2 \cosh ( \beta\mu)\right) + \cdots \;,
\end{align}
where we use $
\cos (i \beta\mu + \frac{k}{n} \frac{\ell_t}{L}) 
=  \cosh (\beta\mu) - i \frac{k}{n} \frac{\ell_t}{L} \sinh (\beta\mu)  -\frac{1}{2} ( \frac{k}{n} \frac{\ell_t}{L})^2 \cosh (\beta\mu) + \cdots \;.
$
Then
\begin{align}\label{EELargeSpaceLimitRSector}
S_n^\mu &= \frac{1}{1- n} \sum_{k=-\frac{n-1}{2}}^{\frac{n-1}{2}} \log \Big| \frac{\cos(\pi z_1)}{\cos(\pi z_2)} \frac{\prod_{m=1}^\infty (1 + q^{2m} + 2 \cos (2\pi z_1) q^m)}{ \prod_{m=1}^\infty (1 + q^{2m} + 2 \cos (2\pi z_2) q^m)} \Big|^2\nonumber \\
&= \frac{1}{1- n} \sum_{k=-\frac{n-1}{2}}^{\frac{n-1}{2}}  \left\{ \log \Big|
1 - i \frac{k}{n} \frac{\ell_t}{2L} \tanh (\frac{\beta\mu}{2})  -\frac{1}{8} (\frac{k}{n} \frac{\ell_t}{L})^2 + \cdots \Big|^2 \right. \nonumber \\
&\qquad\qquad\qquad\qquad  \left.  + \sum_{m=1}^{\infty} \log \Big| 1 - \frac{i \frac{k}{n} \frac{\ell_t}{L} \sinh (\beta\mu) + \frac{1}{2} (\frac{k}{n} \frac{\ell_t}{L})^2 \cosh (\beta\mu)}{\cosh (\beta m) + \cosh (\beta \mu) } + \cdots \Big|^2 \right\}  \nonumber \\
&=\frac{-1}{1\- n}\! \sum_{k=-\frac{n-1}{2}}^{\frac{n-1}{2}} \! \left(\frac{k}{n} \frac{\ell_t}{2L} \right)^2 \Big[ 1 + \tanh^2 (\frac{\beta\mu}{2}) 
+ \sum_{m=1}^{\infty} \frac{ 4 + 4 \cosh (\beta\mu) \cosh (\beta m)}{(\cosh (\beta m) + \cosh (\beta \mu) )^2 } \Big]  \!+\! \cdots\nonumber  \\ 
&= \frac{n+1}{12 n} \frac{\ell_t^2}{4L^2}  \left[ 1 + \tanh^2 (\frac{\beta\mu}{2}) 
+ \sum_{m=1}^{\infty} \frac{ 4 + 4 \cosh (\beta\mu) \cosh (\beta m)}{(\cosh (\beta m) + \cosh (\beta \mu) )^2 }\right]  + \mathcal O \left(\frac{\ell_t}{L} \right)^4 \;.
\end{align} 
This result is valid for $\frac{\ell_t}{L} \ll 1 $.  
By taking $n\to 1$, we get 
\begin{align}
S^\mu 
&= \frac{1}{6} \frac{\ell_t^2}{4L^2} \left[ 1 + \tanh^2 (\frac{\beta\mu}{2}) 
+ \sum_{m=1}^{\infty} \frac{ 4 + 4 \cosh (\beta\mu) \cosh (\beta m)}{(\cosh (\beta m) + \cosh (\beta \mu) )^2 }\right]  + \mathcal O \left(\frac{\ell_t}{L} \right)^4 \;.
\end{align} 
Thus we have demonstrated that the entanglement and R\'enyi entropies vanish as fast as $ \left(\frac{\ell_t}{L} \right)^2$ for the large radius limit, $\frac{\ell_t}{L} \ll 1 $. 

We also demonstrate this for the Dirac fermion in the NS-sector with an anti-periodic boundary condition. The entanglement entropy is given by \eqref{SAmu3}
\begin{align}
S^\mu &= \frac{1}{1- n} \sum_{k=-\frac{n-1}{2}}^{\frac{n-1}{2}} \log \Big|\frac{\vartheta_3 (\frac{k}{n} \frac{\ell_t}{2\pi L}+\frac{i\beta\mu}{2\pi} |i\beta )}{\vartheta_3 (\frac{i\beta\mu}{2\pi} |i\beta)} \Big|^2 \bigg|_{n=1} \;.
\end{align} 
A slightly modified representation for \eqref{Theta3} is useful as  $
\vartheta_3 (z|\tau) = \prod_{m=1}^\infty (1 - q^m)(1 + q^{2m-1} + 2 \cos (2\pi z) q^{m-1/2}) $
along with the notations  $y_1 = e^{- \beta\mu + i \frac{k}{n} \frac{\ell_t}{L}}, ~y_2 = e^{-\beta\mu} , ~ q=e^{-\beta}$. Thus
\begin{align}\label{EELargeSpaceLimitNSSector}
S^\mu &= \frac{2}{3} \frac{\ell_t^2}{4L^2} \sum_{m=1}^{\infty} \frac{ 1 + \cosh (\beta\mu) \cosh ((m-1/2) \beta )}{(\cosh ((m-1/2) \beta ) + \cosh (\beta \mu) )^2 }  + \mathcal O \left(\frac{\ell_t}{L} \right)^4 \;, 
\end{align} 
where we again use a series expansion for $\frac{\ell_t}{L} \ll 1 $. Thus the entanglement entropy vanishes at least as $ \ell_t^2/L^2 $ as the size approaches infinite space limit. These results confirm the earlier claim \cite{CardySlides:2016} and extend the results for the multi-interval case in a straightforward manner.

\subsubsection{High temperature limit} 

The high temperature limit of the entropies shows quite different behavior compared to that of low temperature. It turns out that the limit is sensitive to the value of $\alpha$ due to various arguments proportional to $\frac{1}{\tau}$ in the $\vartheta$ functions. In the high temperature limit $\beta \to 0$, we get 
\begin{align}\label{DifferentLimitsHighTT}
\frac{1}{\tau} = \frac{1}{\tau_1 + i \tau_2} = \frac{2\pi }{\alpha + i \beta}  \rightarrow \left\{ \begin{array}{lll}
-i \frac{2\pi }{ \beta} \;,  & \qquad  & \alpha=0 \;, \\  
\frac{2\pi }{\alpha }\;,  & & \alpha\neq 0 \;. \end{array} \right.  
\end{align} 
Thus we explorer these two different cases separately in general. Due to the reason we focus on $J=\alpha=0$ in this section. We consider this case first and comment about the other later. 

\paragraph{$\alpha=0$ case:} 
Using \eqref{GeneralEEFormulaJmuHighT} and focusing on the case $\alpha=2\pi \tau_1=J=0$, one can see that the entropies in the NSNS sector, with anti-periodic conditions both on the spatial and temporal circles, reduces to the formula 
\begin{align}
\tilde S_n^{\mu}
= \frac{1}{1- n} \sum_{k=-\frac{n-1}{2}}^{\frac{n-1}{2}} \log \Big|\frac{e^{-\frac{2\pi^2}{\beta} (\frac{k \ell_t}{n 2\pi L}+ i \frac{\beta \mu}{2\pi})^2}}{e^{\frac{\beta \mu^2}{2}}} \frac{\vartheta_3(\mu - \frac{i}{\beta}\frac{k}{n} \frac{\ell_t}{L}|\frac{2\pi i}{\beta})}{\vartheta_3(\mu|\frac{2\pi i}{\beta})} \Big|^2   \;.
\end{align} 
Here we note that the high temperature limit for the NSNS sector has the same as $\vartheta_3$ that is used in the low temperature limit.  

Using the product representation in \eqref{Theta3}, we get the R\'enyi entropy as 
\begin{align}\label{RenyiSAmuNSHighT}
\tilde S_n^\mu &= \frac{1}{1- n}  \sum_{k=-\frac{n-1}{2}}^{\frac{n-1}{2}} \log \Big|\frac{e^{-\frac{2\pi^2}{\beta} (\frac{k \ell_t}{n 2\pi L}+ i \frac{\beta \mu}{2\pi})^2}}{e^{\frac{\beta \mu^2}{2}}} \times \prod_{m=1}^{\infty} \frac{(1 - q^m)(1 + y_1 q^{m-1/2})(1 + y_1^{-1} q^{m-1/2})}{(1 - q^m)(1 + y_2 q^{m-1/2})(1 + y_2^{-1} q^{m-1/2})} \Big|^2  \nonumber \\
&=\frac{n+1}{3n} \frac{1}{\beta} \frac{\ell_t^2}{4L^2} + 2\sum_{l=1}^{\infty} \frac{(-1)^{l-1}}{l} 
\frac{\cos (2\pi l\mu)}{\sinh (\pi l/\beta )} \frac{1}{1-n}\left[\frac{\sinh(\frac{\pi \ell_t }{\beta L}l)}{\sinh(\frac{2\pi^2 \ell_t }{n \beta L}l)} -n \right]  
\;, 
\end{align} 
where we use the following identifications 
$y_1 = e^{2\pi i \mu + \frac{2\pi}{\beta} \frac{k}{n} \frac{\ell_t}{L}}$, $y_2 = e^{2\pi i \mu}$, and  $q=e^{-\frac{4\pi^2}{\beta}}$. Here we use the high temperature expansion, $\beta \to 0$. The computation is similar to that in the low temperature limit. 

By taking $n\to 1$ limit, we get the entanglement entropy in the high temperature limit, 
\begin{align}\label{SAmuNSHighT}
\tilde S^\mu &= \frac{2}{3\beta} \frac{\ell_t^2}{4L^2} +  2\sum_{l=1}^{\infty} \frac{(-1)^{l-1}}{l} 
\frac{\cos (2\pi l\mu)}{\sinh (2\pi^2 l/\beta )} \left[1- \frac{\pi l}{\beta} \frac{\ell_t}{L} \cot \left( \frac{\pi l}{\beta} \frac{\ell_t}{L} \right) \right]  
\;. 
\end{align} 
Here we comment on the results \eqref{RenyiSAmuNSHighT} and \eqref{SAmuNSHighT}. We focus on the chemical potential dependent part of the R\'enyi and entanglement entropies, and these results are explicitly evaluated in \cite{Ogawa:2011bz}\cite{Herzog:2013py}. If we set $\mu \to 0$, the result reduces to that given in \cite{Azeyanagi:2007bj}, which also points out that the total entanglement entropy (including the chemical potential independent part) is the same as the thermal entropy. Note that the properties in the high temperature limit is quite different because there is no non-trivial limit that facilitates the interplay between the chemical potential and the energy levels of the Dirac fermion on a circle. 

We also consider the high temperature limit of the R-sector. The general formula \eqref{GeneralEEFormulaJmuHighT} in the RNS sector, with a periodic boundary condition on the spatial circle, reduces to 
\begin{align}
\tilde S_n^{\mu}
= \frac{1}{1- n} \sum_{k=-\frac{n-1}{2}}^{\frac{n-1}{2}} \log \Big|\frac{e^{-\frac{2\pi^2}{\beta} (\frac{k \ell_t}{n 2\pi L}+ i \frac{\beta \mu}{2\pi})^2}}{e^{\frac{\beta \mu^2}{2}}} \frac{\vartheta_4(\mu - \frac{i}{\beta}\frac{k}{n} \frac{\ell_t}{L}|\frac{2\pi i}{\beta})}{\vartheta_4(\mu|\frac{2\pi i}{\beta})} \Big|^2   \;.
\end{align} 
We note that the differences between $\vartheta_3$ given in \eqref{Theta3} and $\vartheta_4$ are a couple of signs. 
\begin{align}\label{Theta4}
\vartheta_4 (z|\tau) &= \prod_{m=1}^\infty (1 - q^m)(1 - y q^{m-1/2})(1 - y^{-1} q^{m-1/2}) \;.
\end{align} 
The entropies are similar to those of the NSNS sector. For example, the entanglement entropy is 
\begin{align}\label{SAmuNSHighTR}
\tilde S^\mu &= \frac{2}{3\beta} \frac{\ell_t^2}{4L^2} -  2\sum_{l=1}^{\infty} \frac{1}{l} 
\frac{\cos (2\pi l\mu)}{\sinh (2\pi^2 l/\beta )} \left[1- \frac{\pi l}{\beta} \frac{\ell_t}{L} \cot \left( \frac{\pi l}{\beta} \frac{\ell_t}{L} \right) \right]  
\;. 
\end{align} 
There is a relative sign between the two terms that comes from the sign of $\vartheta_4$. Thermal entropy is also sensitive to the sign as mentioned in \cite{Azeyanagi:2007bj}.   

\paragraph{$\alpha\neq 0$ case:}
Before moving on, we would like to mention that \eqref{SAmuNSHighT} and \eqref{SAmuNSHighTR} are not the only possible behaviors of the entropies in the high temperature limit, while they are directly related to the known thermal entropies. Recall the two limits given in \eqref{DifferentLimitsHighTT}. We compute the entropies for $\alpha\neq 0$ case in the high temperature limit, $\beta\to 0$. 

Consider \eqref{GeneralEEFormulaJmuHighT} and $J=0$. Now in the presence of non-zero $\alpha=2\pi \tau_1$, we can set $\beta\to 0$ to compute the dominant contribution. The NSNS and RNS sectors are similar for the high temperature limit, we consider them together. 
\begin{align}\label{SAmuNSHighTRAlphaZero1}
\tilde S_n^{\mu}
&= \frac{1}{1- n} \sum_{k=-\frac{n-1}{2}}^{\frac{n-1}{2}} \log \Big|e^{-i\frac{1}{2\alpha} (\frac{k \ell_t}{nL})^2} \frac{\vartheta_{3,4}(\frac{k}{n} \frac{\ell_t}{L} \frac{1}{\alpha}|-\frac{2\pi}{\alpha})}{\vartheta_{3,4}(0|-\frac{2\pi}{\alpha})} \Big|^2 \nonumber \\
&= \frac{2}{1 \- n} \sum_{m=1}^{\infty} \sum_{k=-\frac{n-1}{2}}^{\frac{n-1}{2}} \log \Big[\frac{1 \pm \cos (\frac{2\pi}{\alpha} [m-\frac{1}{2}] - \frac{1}{\alpha} \frac{k}{n} \frac{\ell_t}{L}) }{1 \pm \cos (\frac{2\pi}{\alpha} [m-\frac{1}{2}])} \Big]
\;. 
\end{align} 
Where the indices $3,+$ and $4,-$ indicate NSNS and RNS sectors, respectively. Even though we are not able to sum over the index $k$ explicitly, it is clear to see that the R\'enyi entropy is oscillating function of $\alpha$ and the sub-system size $\ell_t$.

\subsection{Current} \label{sec:EEJ}

To appreciate the qualitative effects of the current $J$, it is useful to consider the twisted boundary condition \eqref{TwistedeBC} as well. The mode expansion of the Dirac fermion has the form 
\begin{align}
\psi_- =  \sum_{\tilde r \in \mathbb Z + a + J} \psi_r (t) e^{i \tilde r s} \;.
\end{align}
It is clear that the presence of current changes the periodicity of the compact fermion. Note that one can change the periodic fermion into the anti-periodic fermion by changing the strength of the current $J$. For $a=0$, the fermion is periodic with $J=0$, while it is anti-periodic with $J=1/2$. One can expect that this would produce physical effects on the entropies. 

Even before performing any more computations with current, we can appreciate the physical effects of the current based on the computations done already in the previous sections. Consider one of the simplest cases for $\alpha=2\pi \tau_1=0$ and $\mu=0$ at the zero temperature limit $T\to 0$. Let us choose $a=0$ and increase the current from $J=0$ to $J=1/2$ for the fixed anti-periodic boundary condition for the temporal circle. This changes the boundary condition in the spatial circle from the periodic one to the anti-periodic one. The corresponding entanglement entropies have been already computed in \eqref{SAmuNS} and \eqref{SAmuR}. By setting $\mu =0$, we get 
\begin{align}\label{EECurrentFromMu}
S = \left\{ \begin{array}{lll}
2\sum_{l=1}^{\infty} \frac{(-1)^{l-1}}{l} \frac{1}{\sinh ( l\beta/2 )} \left[1- l \frac{\ell_t}{2L} \cot \left(l \frac{\ell_t}{2L} \right) \right]\;,  & \qquad  & J=0 \;, \\  & & \\
2\sum_{l=1}^{\infty} \frac{(-1)^{l-1}}{l} \frac{\cosh ( l\beta/2 )}{\sinh ( l\beta/2 )} \left[1- l \frac{\ell_t}{2L} \cot \left(l \frac{\ell_t}{2L} \right) \right]\;,  & & J=1/2 \;. \end{array} \right.  
\end{align} 
At zero temperature, the entanglement entropy vanishes for $J=0$, while it has a non-zero contribution for $J=1/2$. Thus dialing current brings visible effects in the entanglement entropy. It will be interesting to consider some experimental realizations of these effects. 

One can consider more general boundary conditions by including both $a$ and $J$. We are going to study $a+J=0$ and $a+J=1/2$ separately below. We postpone the study for the more general boundary conditions for $0<a+J<1/2$.  

\subsubsection{Anti-periodic fermion}

Let us consider first the anti-periodic fermion, that satisfies $a+J=1/2$. From \eqref{GeneralEEFormulaJmu}, one can get the R\'enyi entropy formula with the current dependence  
\begin{align}\label{SAmu3J}
S_n^J &= \frac{1}{1 \- n}\sum_{k=-\frac{n-1}{2}}^{\frac{n-1}{2}} \log \Big|\frac{\vartheta_3 (\frac{k}{n} \frac{\ell_t}{2\pi L}+\tau_1 J |\tau )}{\vartheta_3 (\tau_1 J |\tau)} \Big|^2  \;, 
\end{align} 
where we set $\mu =0$ for simplicity. We study the entropies as a function of the current $J$ for fixed $a+J=1/2$ in the low temperature limit, in the large radius limit, and in the high temperature limit in turn. The computations are brief because they are similar to the previous sections.

\paragraph{Low temperature limit:} 
We compute the entropies \eqref{SAmu3J}, in the low temperature limit $\beta \to \infty$, by using the product representation \eqref{Theta3}.
The details are similar to those in \eqref{RenyiSAmuNS} by using the notations $y_1 = e^{i \alpha J + i \frac{k}{n} \frac{\ell_t}{L}}, ~y_2 = e^{i \alpha J}, ~ q=e^{2\pi i \tau} = e^{i\alpha -\beta}$. The R\'enyi entropy reads 
\begin{align}\label{RenyiSAmuNSJ}
S_n^J 
&= 4 \sum_{l,m=1}^{\infty} \frac{(-1)^{l-1}}{l}  \frac{\cos ([m-1/2]\alpha l)~ \cos (\alpha J l)}{e^{(m-1/2) \beta l}} \frac{1}{1\- n} \left[   \sin \left(\frac{\ell_t}{2L}l\right) \csc \left(\frac{1}{n} \frac{\ell_t}{2L}l\right) \-n\right]
\;. 
\end{align} 
The corresponding entanglement entropy is 
\begin{align}\label{SAmuNSJ}
S^J 
&= 4 \sum_{l,m=1}^{\infty} \frac{(-1)^{l-1}}{l}  \frac{\cos ([m-1/2]\alpha l) ~\cos (\alpha J l)}{e^{(m-1/2) \beta l}}
 \left[1- l \frac{\ell_t}{2L} \cot \left(l \frac{\ell_t}{2L} \right) \right]  
\;. 
\end{align} 
The entropies depend on the current $J$ non-trivially. That is markedly different from the dependence of the entropies of the chemical potential $\mu$. Furthermore it is more interesting to note that the entropies depend on the parameter $\alpha = 2\pi \tau_1$ in two different ways with two cosine functions. 

To see the effects of $J$ and $\alpha$ more clearly, we consider the case $l=m=1$ which provides the dominant contribution to the entropies in the low temperature limit.
\begin{align}
4 e^{-\frac{\beta}{2}}  \cos (\alpha/2)
\cos (\alpha J) \left[1- \frac{\ell_t}{2L} \cot \left(\frac{\ell_t}{2L} \right) \right]  
\;. 
\end{align} 
For a fixed $\alpha = 2\pi \tau_1$, the change of current produces an oscillating behavior of the entropies by the cosine function. When we dial the parameter $\alpha$ for a fixed $J$, the product of two cosine functions produce an `interference pattern.' The interference pattern would produce the following beat frequency $f_b$ depending on the strength of the current as 
\[ \cos (\alpha/2) \cos (\alpha J) \longrightarrow \left\{ \begin{array}{lll}
f_b = J/\pi\;,  & \qquad  & J<1/2 \;, \\  & & \\
f_b = 1/2\pi\;,  & & J>1/2 \;. \end{array} \right. \] 
The beat frequency is easy to measure! In this way one can identify the current $J$ and the dependence of $J$ in the entropies.  
 
\paragraph{Large Radius limit:} 
In turn, we compute the entropies \eqref{SAmu3J} in the large radius limit, $\ell_t/L \to 0$. We use the modified product representation \eqref{Theta3} that is written around the equation \eqref{EELargeSpaceLimitNSSector} along with the notation $y_1 = e^{i \alpha J + i \frac{k}{n} \frac{\ell_t}{L}}, ~y_2 = e^{i \alpha J}, ~ q=e^{2\pi i \tau} \equiv e^{i\alpha -\beta}$.   
\begin{align}\label{SAmuNSJLargeL}
S_n^J 
&= \frac{(n+1)}{3n} \frac{\ell_t^2}{4L^2} \sum_{m=1}^\infty \left[ 
\frac{[\cosh ([m\-\frac{1}{2}]\beta) \cos ([m\-\frac{1}{2}]\alpha) \+ 1]  \cos (\alpha J)}{[\cosh ([m\-\frac{1}{2}]\beta) \cos ([m\-\frac{1}{2}]\alpha) \+   \cos (\alpha J)]^2 \+ \sinh^2 ([m\-\frac{1}{2}]\beta) \sin^2 ([m\-\frac{1}{2}]\alpha)} \right. \nonumber \\
&\qquad \left. + 
\frac{([\cosh ([m\-\frac{1}{2}]\beta) \cos ([m\-\frac{1}{2}]\alpha) \+ \cos (\alpha J)]^2 \- \sinh^2 ([m\-\frac{1}{2}]\beta) \sin^2 ([m\-\frac{1}{2}]\alpha)  )  \sin^2 (\alpha J)}{([\cosh ([m\-\frac{1}{2}]\beta) \cos ([m\-\frac{1}{2}]\alpha) \+   \cos (\alpha J)]^2 \+ \sinh^2 ([m\-\frac{1}{2}]\beta) \sin^2 ([m\-\frac{1}{2}]\alpha) )^2} \right] \nonumber \\
&\qquad + \mathcal O(\frac{\ell_t^2}{L^2})^4 \;. 
\end{align} 
Taking $n\to 1$ limit is straightforward. While the result is a little bit messy, it is clear to check that the entropies vanish at least $\mathcal O(\frac{\ell_t^2}{L^2})^2$ as we take the large radius limit ${\ell_t}/{L} \to 0 $. The result extends the earlier claim \cite{CardySlides:2016} in the presence of current. It is straightforward to generalize the result for the multi-interval case. 

\paragraph{High temperature limit:} 
We compute the entropies \eqref{GeneralEEFormulaJmuHighT} that is valid for the high temperature limit $\beta \to 0$. We set $a+J=1/2$ and $\mu=0$ for simplicity. Then we get 
\begin{align}
\tilde S_n^{J}
= \frac{1}{1 \- n} \sum_{k=-\frac{n-1}{2}}^{\frac{n-1}{2}} \log \Big|\frac{e^{-i\frac{\pi}{\tau} (\frac{k \ell_t}{n2\pi L}+ \tau_1 J )^2}}{e^{-i\frac{\pi}{\tau} ( \tau_1 J )^2}} \frac{\vartheta_3(\frac{k}{n} \frac{\ell_t}{2\pi L} \frac{1}{\tau} + \frac{\tau_1 J}{\tau}|-\frac{1}{\tau})}{\vartheta_3(\frac{\tau_1 J }{\tau}|-\frac{1}{\tau})} \Big|^2  \;.
\end{align} 
The limit is sensitive to the presence of $\alpha$. Thus we study two cases $\alpha= 0$ and $\alpha\neq 0$ in turn. 

For $\alpha=2\pi \tau_1= 0$, we set $\tau= i\beta/2\pi$. Thus 
\begin{align}\label{RenyiSAmuNSJHighTCurrent1}
\tilde S_n^{J}
&= \frac{1}{1 \- n} \sum_{k=-\frac{n-1}{2}}^{\frac{n-1}{2}} \log \Big|e^{-\frac{1}{2\beta} (\frac{k \ell_t}{nL})^2} \frac{\vartheta_3(-i\frac{k}{n} \frac{\ell_t}{L} \frac{1}{\beta}|i\frac{2\pi}{\beta})}{\vartheta_3(0|i\frac{2\pi}{\beta})} \Big|^2 \nonumber \\
&= \frac{(n+1)}{3n \beta} \frac{\ell_t^2}{4L^2} 
+ \sum_{l=1}^\infty \frac{(-1)^{l-1}}{l} \frac{2}{\sinh (\frac{\pi}{\beta}l)} 
\frac{1}{1\- n} \left[   \sinh \left(\frac{\pi}{\beta} \frac{\ell_t}{L}l\right) \text{csch} \left(\frac{2\pi^2}{\beta n} \frac{\ell_t}{L}l\right) \-n\right] \;.
\end{align} 
It is straightforward to get the corresponding entanglement entropy by taking $n \to 1$ limit. The first term needs to be included to $\tilde S^0$ to see the full entropies. The second term depends on $J$ through the boundary condition $a+J=1/2$ and thus the Hilbert space.

Now we turn to $\alpha\neq 0$ case. 
Due to the presence of $\alpha$, one can take the zero temperature limit in a straightforward manner as $ \frac{1}{\tau} = \frac{2\pi}{\alpha + i \beta} = \frac{2\pi (\alpha -i \beta)}{\alpha^2 + \beta^2} \to \frac{2\pi}{\alpha} $. 
Thus
\begin{align}
\tilde S_n^{J}
= \frac{1}{1 \- n} \sum_{k=-\frac{n-1}{2}}^{\frac{n-1}{2}} \log \Big|\frac{e^{-i\frac{2\pi^2}{\alpha} (\frac{k \ell_t}{n 2\pi L}+ \frac{\alpha}{2\pi} J )^2}}{e^{-i\frac{2\pi^2}{\alpha} ( \frac{\alpha}{2\pi} J )^2}} \frac{\vartheta_3(\frac{k}{n} \frac{\ell_t}{L} \frac{1}{\alpha} + J|-\frac{2\pi}{\alpha})}{\vartheta_3(J|-\frac{2\pi}{\alpha})} \Big|^2  \;.
\end{align} 
The first factor inside the log is pure imaginary and would not contribute. The R\'enyi entropy reads
\begin{align}\label{RenyiSAmuNSJHighT}
\tilde S_n^J =
 \frac{2}{1 \- n} \sum_{m=1}^{\infty} \sum_{k=-\frac{n-1}{2}}^{\frac{n-1}{2}} \log \Big[\frac{\cos (2\pi J) + \cos (\frac{2\pi}{\alpha} [m-\frac{1}{2}] - \frac{2\pi}{\alpha} \frac{k}{n} \frac{\ell_t}{L}) }{\cos (2\pi J) + \cos (\frac{1}{\alpha} [m-\frac{1}{2}])} \Big]
\;. 
\end{align} 
Unfortunately, the sum over $k$ is not easy to evaluate. There are further sub-leading contributions as $\beta \to 0$. It will be interesting to perform a numerical study for this limit.

\subsubsection{Periodic fermion}

Let us consider the periodic fermion that satisfies the condition $a+J=0$. From \eqref{GeneralEEFormulaJmu}, one can get the R\'enyi entropy with the current dependence  
\begin{align}\label{SAmu2J2}
S_n^J &= \frac{1}{1 \- n}\sum_{k=-\frac{n-1}{2}}^{\frac{n-1}{2}} \log \Big|\frac{\vartheta_2 (\frac{k}{n} \frac{\ell_t}{2\pi L}+\tau_1 J |\tau )}{\vartheta_2 (\tau_1 J |\tau)} \Big|^2  \;, 
\end{align} 
where we set $\mu =0$. We study the entropies as a function of the current $J$ in the low temperature limit, the large radius limit, and the high temperature limit in turn for $a+J=0$. 
There are interesting physical differences between the periodic and anti-periodic fermions that come along from the difference between $\vartheta_2$ and $\vartheta_3$. $\vartheta_2$ \eqref{Theta2} has a front factor in addition to the infinite product.   

\paragraph{Low temperature limit:} 
The computation for the entanglement entropy \eqref{SAmu2J2} for $\beta \to \infty$ is similar to that of the anti-periodic fermion done in \eqref{RenyiSAmuNSJ}. The difference is to use the $\vartheta_2$ given in \eqref{Theta2} instead of $\vartheta_3$. Thus  
\begin{align}\label{RenyiSAmuRJ22}
S_n^J
&= 2 \sum_{l=1}^{\infty} \frac{(-1)^{l-1}}{l} \cos (\alpha J l)\left( 1+ \sum_{m=1}^{\infty} 2\frac{\cos (m\alpha l)}{e^{m \beta l}} \right)\frac{1}{1\- n} \left[   \sin \left( \frac{\ell_t}{2L}l\right) \csc \left(\frac{1}{n} \frac{\ell_t}{2L}l\right) \-n\right] 
\;. 
\end{align} 
The entanglement entropy has the form. 
\begin{align}\label{SAmuRJ22}
S^J
&= 2 \sum_{l=1}^{\infty} \frac{(-1)^{l-1}}{l} \cos (\alpha J l)\left( 1+ \sum_{m=1}^{\infty} 2\frac{\cos (m\alpha l)}{e^{m \beta l}} \right)
 \left[1- l \frac{\ell_t}{2L} \cot \left(l \frac{\ell_t}{2L} \right) \right]  
\;. 
\end{align} 
The entropies have a non-trivial dependence on current $J$, only on the combination of $\alpha J$, in the zero temperature limit. The contributions with the sum over $m$ have the similar properties such as beat frequency compared to those of the anti-periodic fermion mentioned in the previous section. 

The first term in the round brackets in \eqref{RenyiSAmuRJ22} and \eqref{SAmuRJ22}, coming from the front factor of the $\vartheta_2$, gives a distinct physical significance. It is actually non-vanishing at zero temperature as we discuss already for the second case in \eqref{EECurrentFromMu}.   
\begin{align}\label{SAmuRJ22ZeroT}
S^J (T=0)
&= 2 \sum_{l=1}^{\infty} \frac{(-1)^{l-1}}{l} \cos (\alpha J l)
 \left[1- l \frac{\ell_t}{2L} \cot \left(l \frac{\ell_t}{2L} \right) \right]  
\;. 
\end{align} 
We note that the entropies oscillate as we change $\alpha J$. Thus they can become negative. Of course, we need to include the entropy $S^0$ that is independent of chemical potential and current. 

\paragraph{Large radius limit:} 
In the large radius limit, $\ell_t/L \to 0$, the entropies \eqref{SAmu2J2} can be evaluated similar to the previous case with the anti-periodic fermion \eqref{SAmuNSJLargeL} to give 
\begin{align}\label{SAmuRJLargeL2}
S_A^J 
&= \frac{2}{3} \frac{\ell_t^2}{4L^2} \Bigg( \frac{\cos^{-2} \left[\frac{\alpha J}{2} \right] }{4}  \+  \Bigg.  \sum_{m=1}^\infty \left[ 
\frac{[\cosh (m\beta) \cos (m\alpha) \+ 1]  \cos (\alpha J)}{[\cosh (m\beta) \cos (m\alpha) \+   \cos (\alpha J)]^2 \+ [\sinh (m\beta) \sin (m\alpha)]^2 } \right. \nonumber \\
&\qquad \left. + 
\frac{([\cosh (m\beta) \cos (m\alpha) \+ \cos (\alpha J)]^2 \- \sinh^2 (m\beta) \sin^2 (m\alpha)  )  \sin^2 (\alpha J)}{([\cosh (m\beta) \cos (m\alpha) \+   \cos (\alpha J)]^2 \+ [\sinh (m\beta) \sin (m\alpha)]^2 )^2} \right] \Bigg. \Bigg)  + \mathcal O(\frac{\ell_t^2}{L^2})^4 \;.  
\end{align} 
Once again, the first term is an additional contribution compared to the anti-periodic fermion given in \eqref{SAmuNSJLargeL}. That comes from the cosine factor in the $\vartheta_2$. Nevertheless, all the contributions vanish as fast as $ \mathcal O(\frac{\ell_t^2}{L^2})^2$.

\paragraph{High temperature limit:} 
We compute the entropies \eqref{GeneralEEFormulaJmuHighT} that is valid for the high temperature limit $\beta \to 0$. We set $a+J=0$ and $\mu=0$ for simplicity. 
The R\'nyi entropy can be computes as  
\begin{align}
\tilde S_n^{J}
= \frac{1}{1 \- n} \sum_{k=-\frac{n-1}{2}}^{\frac{n-1}{2}} \log \Big|\frac{e^{-i\frac{\pi}{\tau} (\frac{k \ell_t}{n 2\pi L}+ \tau_1 J )^2}}{e^{-i\frac{\pi}{\tau} ( \tau_1 J )^2}} \frac{\vartheta_4(\frac{k}{n} \frac{\ell_t}{2\pi L} \frac{1}{\tau} + \frac{\tau_1 J}{\tau}|-\frac{1}{\tau})}{\vartheta_4(\frac{\tau_1 J }{\tau}|-\frac{1}{\tau})} \Big|^2  \;.
\end{align} 
Similar to the anti-periodic case, there are two different non-trivial limits depending on $\alpha$. 

For $\alpha=2\pi \tau_1= 0$, we set $\tau= i\beta/2\pi$. Thus 
\begin{align}\label{RenyiHighTCurrent12}
\tilde S_n^{J}
&= \frac{1}{1 \- n} \sum_{k=-\frac{n-1}{2}}^{\frac{n-1}{2}} \log \Big|e^{-\frac{1}{2\beta} (\frac{k \ell_t}{nL})^2} \frac{\vartheta_4(-i\frac{k}{n} \frac{\ell_t}{L} \frac{1}{\beta}|i\frac{2\pi}{\beta})}{\vartheta_4(0|i\frac{2\pi}{\beta})} \Big|^2 \nonumber \\
&= \frac{n+1}{3n \beta} \frac{\ell_t^2}{4L^2} 
- \sum_{l=1}^\infty \frac{1}{l} \frac{2}{\sinh (\frac{2\pi^2}{\beta}l)} 
\frac{1}{1\- n} \left[   \sinh \left(\frac{2\pi^2}{\beta} \frac{\ell_t}{L}l\right) \text{csch} \left(\frac{2\pi^2}{\beta n} \frac{\ell_t}{L}l\right) \-n\right] \;.
\end{align} 
This is similar to that of the NSNS sector. 
It is straightforward to get the corresponding entanglement entropy by taking $n \to 1$ limit. 

Now we turn to $\alpha\neq 0$ case. The R\'nyi entropy has the form
\begin{align}
\tilde S_n^{J}
= \frac{1}{1 \- n} \sum_{k=-\frac{n-1}{2}}^{\frac{n-1}{2}} \log \Big|\frac{e^{-i\frac{2\pi^2}{\alpha} (\frac{k \ell_t}{n 2\pi L}+ \frac{\alpha}{2\pi} J )^2}}{e^{-i\frac{2\pi^2}{\alpha} ( \frac{\alpha}{2\pi} J )^2}} \frac{\vartheta_4(\frac{k}{n} \frac{\ell_t}{L} \frac{1}{\alpha} + J|-\frac{2\pi}{\alpha})}{\vartheta_4(J|-\frac{2\pi}{\alpha})} \Big|^2  \;.
\end{align} 
Similar to the NSNS sector, the first factor inside the log is pure imaginary and would not contribute.  Due to the presence of $\alpha$, one can take the zero temperature limit in a straightforward manner as $ \frac{1}{\tau}  \to \frac{2\pi}{\alpha} $ similar to the previous case. The R\'enyi entropy reads
\begin{align}\label{RenyiSAmuRJHighT}
\tilde S_n^J =
 \frac{2}{1 \- n} \sum_{m=1}^{\infty} \sum_{k=-\frac{n-1}{2}}^{\frac{n-1}{2}} \log \Big[\frac{\cos (2\pi J) - \cos (\frac{2\pi}{\alpha} [m-\frac{1}{2}] - \frac{1}{\alpha} \frac{k}{n} \frac{\ell_t}{L}) }{\cos (2\pi J) - \cos (\frac{2\pi}{\alpha} [m-\frac{1}{2}])} \Big]
\;. 
\end{align} 
The result is similar to that of the anti-periodic fermion given in \eqref{RenyiSAmuNSJHighT}. Only difference is the relative sign between two cosine terms in the log. While the sum over $k$ is not easy to evaluate, it is clear that the individual term contributes with the periodic behavior. There are further sub-leading contributions as $\beta \to 0$. It will be interesting to perform a numerical study for this. \\

\subsection{Chemical potential and Current}\label{sec:EEMuJGeneral}

In this section we consider both the current and chemical potential to compute the entropies for the anti-periodic and periodic fermions. We are going to check that the results reduce to the previous ones in the appropriate limits.   

\subsubsection{Anti-periodic fermion}

Let us consider the anti-periodic fermion first, that has the condition $a+J=1/2$. From \eqref{GeneralEEFormulaJmu}, one can get the entropies with the current and chemical potential dependences  
\begin{align}
S_n^{\mu,J} &= \frac{1}{1\- n} \sum_{k=-\frac{n-1}{2}}^{\frac{n-1}{2}} \log \Big|\frac{\vartheta_3 (\frac{k}{n} \frac{\ell_t}{2\pi L}+\tau_1 J + i \tau_2 \mu |\tau )}{\vartheta_3 (\tau_1 J + i \tau_2 \mu |\tau)} \Big|^2  \;. 
\end{align} 
The computations are straightforward and we quote the results for the appropriate limits. 

\paragraph{Low temperature limit:} 
For the low temperature limit, $\beta \to \infty$, the R\'enyi entropy is  
\begin{align}\label{RenyiEELowTAntiNS}
S_n^{\mu J} 
&= 2 \sum_{l,m=1}^{\infty} \frac{(-1)^{l-1}}{l ~e^{(m-1/2) \beta l}}  \left[ e^{l \beta\mu} \cos (\alpha l (J-m+1/2)) + e^{-l\beta \mu} \cos (\alpha l(J+m-1/2)) \right]\nonumber \\
&\qquad\qquad\qquad \times \frac{1}{1\- n} \left[   \sin \left(\frac{\ell_t}{2L}l\right) \csc \left(\frac{1}{n} \frac{\ell_t}{2L}l\right) \-n\right]  \;, 
\end{align} 
where we use the product representation \eqref{Theta3} along with the notation $y_1 = e^{-\beta \mu + i \alpha J + i \frac{k}{n} \frac{\ell_t}{L}}, ~y_2 = e^{- \beta \mu + i \alpha J}, ~ q=e^{2\pi i \tau} \equiv e^{i\alpha -\beta}$ to follow the previous computations.  
The entanglement entropy is give by taking $n\to 1$ limit 
\begin{align}\label{EELowTAntiNS}
S^{\mu J} 
&= 2\!\! \sum_{l,m=1}^{\infty} \!\!\! \frac{(-1)^{l-1}}{l} \frac{ \cos (\alpha l [J\!\-m\+\frac{1}{2}])}{e^{ (m-\frac{1}{2}-\mu) \beta l}} \! \left[ 1  \+ e^{-2l\beta \mu} \frac{\cos (\alpha l[J\+m\-\frac{1}{2}])}{\cos (\alpha l [J\-m\+\frac{1}{2}])} \right]\!\! \left[1\-  l \frac{\ell_t}{2L} \cot \left(\! l \frac{\ell_t}{2L} \right) \right]. 
\end{align} 
For $J=0$ and $\alpha=0$, the result reduces to \eqref{SAmuNS}. For $\mu=0$, it is also consistent with \eqref{SAmuNSJ} after a small algebra. The result is valid for $-1/2 < \mu <1/2 $. For other values of $\mu$, one can recompute the entropies similar to that of \eqref{EEFinteAtZeroT}. This confirms that the entropies at zero temperature has a non-trivial chemical potential dependence when it coincides with one of the energy levels of the theory.  

\paragraph{Large radius limit:} 
In the large radius limit $\ell_t/L \ll 1$, the computation is rather involved, and thus we provide more details here. We use the product representation \eqref{Theta3} with $y_1 = e^{-\beta \mu + i \alpha J + i \frac{k}{n} \frac{\ell_t}{L}}, ~y_2 = e^{- \beta \mu + i \alpha J}, ~ q=e^{i\alpha -\beta}$.  The key element comes from the following observation 
\begin{align}\label{Llimitqq}
&\frac{(1 + y_1 q^{m-1/2})(1 + y_1^{-1} q^{m-1/2})}{(1 + y_2 q^{m-1/2})(1 + y_2^{-1} q^{m-1/2})} 
= \frac{\cos([m-1/2][\alpha + i \beta]) + \cos(\alpha J + i \beta \mu + \frac{k}{n} \frac{\ell_t}{L})}{\cos([m-1/2][\alpha + i \beta]) + \cos(\alpha J + i \beta \mu)} \nonumber \\
&= 1 - \frac{ \frac{k}{n} \frac{\ell_t}{L} \sin (\alpha J + i \beta \mu) + \frac{1}{2} \frac{k^2}{n^2} \frac{\ell_t^2}{L^2} \cos (\alpha J + i \beta \mu)}{\cos([m-1/2][\alpha + i \beta]) + \cos(\alpha J + i \beta \mu)} + \mathcal O (\frac{\ell_t}{L})^3 \;.
\end{align}
This is an argument of $\log$ and has the following expansion $\log (1+ x) = \sum_{l=1}^\infty \frac{(-1)^{l-1}}{l} x^l$. The linear term proportional to $\ell_t/L$ vanishes with the sum over $k$, and thus the first non-trivial order starts with $(\ell_t/L)^2$. After adding the complex conjugate part and using $\frac{1}{1\- n} [ \sum_{k=-\frac{n-1}{2}}^{\frac{n-1}{2}} \frac{k^2}{n^2} ]  = -(n+1)/(12 n) $, we have 
\begin{align}\label{SAGeneralNSLargeL}
S_n^J 
= \frac{(n+1)}{12 n} \frac{\ell_t^2}{L^2} \sum_{m=1}^\infty & \left[ 
\frac{D_1 \times \cosh (\beta \mu)  \cos (\alpha J) + D_2 \times \sinh (\beta \mu)  \sin (\alpha J)}{D_1^2 + D_2^2} \right. \nonumber \\
&\left. + 
\frac{N_1 \times (D_1^2 - D_2^2) - N_2 \times D_1 \times D_2}{(D_1^2 + D_2^2)^2} \right]  + \mathcal O(\frac{\ell_t^2}{L^2})^4 \;. 
\end{align} 
where 
\begin{align}
D_1 &= \cosh ([m\-1/2]\beta) \cos ([m\-1/2]\alpha) + \cosh(\beta \mu)  \cos (\alpha J) \;,  \nonumber \\
D_2 &= \sinh ([m\-1/2]\beta) \sin ([m\-1/2]\alpha) + \sinh(\beta \mu)  \sin (\alpha J) \;,  \nonumber \\
N_1 &= \cosh^2 (\beta \mu)  \sin^2 (\alpha J) - \sinh^2 (\beta \mu)  \cos^2 (\alpha J) 
= [1-\cosh (2\beta \mu) \cos (2\alpha J)]/2 \;,  \nonumber \\
N_2 &= 4\sinh (\beta \mu) \cosh (\beta \mu)  \sin (\alpha J)    \cos (\alpha J)
= \sinh (2\beta \mu) \sin (2\alpha J)\;. 
\end{align}
The entanglement entropy can be obtained by $n\to 1$. 
Now one can check that this reduces to the case with current ($\mu=0$) evaluated in \eqref{SAmuNSJLargeL}, not to mention to the case with only chemical potential ($\alpha = J=0$) evaluated in \eqref{EELargeSpaceLimitNSSector}. 

Thus the entropies vanish at least as $ \ell_t^2/L^2 $ as the size approaches the infinite space limit. These results confirm the earlier claim for a single interval in the case of the entanglement entropy \cite{CardySlides:2016} and extend to the case with both the chemical potential and current and for the multi-interval case. As we see below, the same is true for the periodic fermion as well.

\paragraph{High temperature limit:}
We compute the entropies \eqref{GeneralEEFormulaJmuHighT} for the anti-periodic boundary condition $a+J=1/2$ in the high temperature limit $\beta \to 0$. We consider the high temperature case with more general setting. As we briefly mention before, the high temperature limit is sensitive to the parameter $\alpha$. For the limit $\beta \to 0$, we get 
\begin{align}
\frac{1}{\tau} = \frac{1}{\tau_1 + i \tau_2} = \frac{2\pi }{\alpha + i \beta}  \rightarrow \left\{ \begin{array}{lll}
-i \frac{2\pi }{ \beta} \;,  & \qquad  & \alpha=0 \;, \\  
\frac{2\pi }{\alpha }\;,  & & \alpha\neq 0 \;. \end{array} \right.  
\end{align} 

We first consider $\alpha=2\pi\tau_1=0$ limit. Then 
\begin{align}
\tilde S_n^{\mu,J}
= \frac{1}{1 \- n} \sum_{k=-\frac{n-1}{2}}^{\frac{n-1}{2}} \log \Big|\frac{e^{-\frac{2\pi^2}{\beta} (\frac{k \ell_t}{n 2\pi L}+ i \frac{\beta \mu}{2\pi})^2}}{e^{-\frac{2\pi^2}{\beta} ( i \frac{\beta \mu}{2\pi})^2}} \frac{\vartheta_3(-i\frac{k}{n} \frac{\ell_t}{L} \frac{1}{\beta} +  \mu|i\frac{2\pi}{\beta})}{\vartheta_3( \mu |i\frac{2\pi}{\beta})} \Big|^2  \;.
\end{align} 
It looks like that the dependence of $J$ disappears as we set $\alpha=0$. The current dependence is still there with the condition $a+J=1/2$.  
The R\'enyi entropy has the following form.
\begin{align}\label{RenyiSAmuNSJHighTGeneral}
\tilde S_n^{\mu,J} 
&\= \frac{(n\+1)}{3n} \frac{1}{\beta} \frac{\ell_t^2}{4L^2} +  \!\sum_{m,l=1}^{\infty} \frac{(-1)^{l-1}\cos [2\pi l \mu ]}{l e^{\frac{4\pi^2 }{\beta}(m-\frac{1}{2})l}} 
   \frac{4}{1\- n}\left[ \sinh \left(\!{\frac{\pi l}{\beta} \frac{ \ell_t}{L}}\!\right) \text{csch} \left(\!{\frac{\pi l}{\beta} \frac{\ell_t}{nL}}\!\right) \- n\right]  \;.
\end{align} 
This result is consistent, after summing over the index $m$, with \eqref{RenyiSAmuNSHighT} and \eqref{RenyiSAmuNSJHighTCurrent1} with $\mu=0$. 

We turn to $\alpha \neq 0$ case. We can get the dominant contribution by taking $\beta\to 0$. Thus 
\begin{align}
\tilde S_n^{\mu,J}
= \frac{1}{1 \- n} \sum_{k=-\frac{n-1}{2}}^{\frac{n-1}{2}} \log \Big|\frac{e^{-i\frac{\pi}{\alpha} (\frac{k \ell_t}{n 2\pi L}+ \frac{\alpha}{2\pi} J )^2}}{e^{-i\frac{\pi}{\alpha} ( \frac{\alpha}{2\pi} J)^2}} \frac{\vartheta_3(\frac{k}{n} \frac{\ell_t}{L} \frac{1}{\alpha} + J|-\frac{2\pi}{\alpha})}{\vartheta_3(J|-\frac{2\pi}{\alpha})} \Big|^2  \;.
\end{align} 
This dominant contribution turns out to be the same as the case with current because the chemical potential dependence disappears with $\beta\to 0$. Thus the R\'enyi entropy reads
\begin{align}\label{RenyiSAmuNSJHighT00}
\tilde S_n^J =
 \frac{2}{1 \- n} \sum_{m=1}^{\infty} \sum_{k=-\frac{n-1}{2}}^{\frac{n-1}{2}} \log \Big[\frac{\cos (2\pi J) + \cos (\frac{2\pi}{\alpha} [m-\frac{1}{2}] - \frac{1}{\alpha} \frac{k}{n} \frac{\ell_t}{L}) }{\cos (2\pi J) + \cos (\frac{2\pi}{\alpha} [m-\frac{1}{2}])} \Big]
\;. 
\end{align} 
This result is consistent with \eqref{SAmuNSHighTRAlphaZero1} with $J=0$ and \eqref{RenyiSAmuNSJHighT}.

\subsubsection{Periodic fermion}

Let us consider the periodic fermion, $a+J=0$. From \eqref{GeneralEEFormulaJmu}, one can get the general formula for the R\'enyi entropy with the chemical potential and current dependence  
\begin{align}
S_n^{\mu,J} &= \frac{1}{1\- n} \sum_{k=-\frac{n-1}{2}}^{\frac{n-1}{2}} \log \Big|\frac{\vartheta_2 (\frac{k}{n} \frac{\ell_t}{2\pi L}+\tau_1 J + i \tau_2 \mu |\tau )}{\vartheta_2 (\tau_1 J + i \tau_2 \mu |\tau)} \Big|^2 \;. 
\end{align} 
Here we also note the $\vartheta_2$ \eqref{Theta2}, compared to $\vartheta_3$, has extra cosine factor that contributes to the entropies.  

\paragraph{Low temperature limit:} 
For the low temperature limit, $\beta \to \infty$, the R\'enyi entropy is 
\begin{align}
S_n^{\mu J} 
&= 2 \sum_{l=1}^{\infty} \!\! \frac{(-1)^{l-1}}{l} e^{ - \mu \beta l} \cos (\alpha J l) \frac{1}{1\- n} \left[   \sin \left(\frac{\ell_t}{2L}l\right) \csc \left(\frac{1}{n} \frac{\ell_t}{2L}l\right) \-n\right]   \\
&+ 2\! \sum_{l,m=1}^{\infty} \!\! \frac{(-1)^{l-1}}{l} \frac{ \cos (\alpha l [J\!\-m])}{e^{ (m -\mu) \beta l}} \! \left[ 1  \+ e^{-2l\beta \mu} \frac{\cos (\alpha l[J\+m])}{\cos (\alpha l [J\-m])} \right]\frac{1}{1\- n} \left[   \sin \left(\frac{\ell_t}{2L}l\right) \csc \left(\frac{1}{n} \frac{\ell_t}{2L}l\right) \-n\right] \;.  \nonumber
\end{align} 
The result is valid for $0 \leq \mu <1 $. We note that the first line gives a non-vanishing contribution for $\mu \beta = const.$ for $\mu \to 0$ and $\beta \to \infty$. One can repeat the computation for $0 < \mu <2 $ with a slightly modified theta function as above. 
And the entanglement entropy is
\begin{align}
S^{\mu J} 
&= 2 \sum_{l=1}^{\infty} \!\! \frac{(-1)^{l-1}}{l} e^{ - \mu \beta l} \cos (\alpha J l) \left[1\- l \frac{\ell_t}{2L} \cot \left(\! l \frac{\ell_t}{2L} \right) \right] \nonumber  \\
&+ 2\! \sum_{l,m=1}^{\infty} \!\! \frac{(-1)^{l-1}}{l} \frac{ \cos (\alpha l [J\!\-m])}{e^{ (m -\mu) \beta l}} \! \left[ 1  \+ e^{-2l\beta \mu} \frac{\cos (\alpha l[J\+m])}{\cos (\alpha l [J\-m])} \right]\!\! \left[1\-l \frac{\ell_t}{2L} \cot \left(\! l \frac{\ell_t}{2L} \right) \right]. 
\end{align} 
Extending the computations to the other values is straightforward and reveals a new result as discussed in the previous sections. For $J=0$ and $\alpha=0$, the result reduces to \eqref{SAmuR}. For $\mu=0$, it is also consistent with \eqref{SAmuRJ22}.

\paragraph{Large radius limit:} 
For the large radius limit $\ell_t/L \ll 1$, the computation is similar to that of the previous section. Thus 
\begin{align}\label{RenyiSAGeneralRLargeL} 
S_n^J 
= \frac{n+1}{12 n} \frac{\ell_t^2}{L^2} & \Bigg(\frac{1}{4}+\frac{1-1/2 (\cosh (2\beta \mu) + \cos (2\alpha J))}{4(\cosh (\beta \mu) + \cos (\alpha J) )^2}  \bigg. \nonumber \\
& \left.+  \sum_{m=1}^\infty  \left[ 
\frac{\tilde D_1 \times \cosh (\beta \mu)  \cos (\alpha J) + \tilde D_2 \times \sinh (\beta \mu)  \sin (\alpha J)}{\tilde D_1^2 + \tilde D_2^2} \right. \right. \nonumber \\
&\qquad\quad \left. + 
\frac{\tilde N_1 \times (\tilde D_1^2 - \tilde D_2^2) - \tilde N_2 \times \tilde D_1 \times \tilde D_2}{(\tilde D_1^2 + \tilde D_2^2)^2} \right]  \bigg.\Bigg) + \mathcal O(\frac{\ell_t^2}{L^2})^4 \;. 
\end{align} 
where 
\begin{align}
\tilde D_1 &= \cosh (m \beta) \cos (m\alpha) + \cosh(\beta \mu)  \cos (\alpha J) \;, \nonumber \\
\tilde D_2 &= \sinh (m\beta) \sin (m\alpha) + \sinh(\beta \mu)  \sin (\alpha J) \;, \nonumber \\
\tilde N_1 &= \cosh^2 (\beta \mu)  \sin^2 (\alpha J) - \sinh^2 (\beta \mu)  \cos^2 (\alpha J) 
= [1-\cosh (2\beta \mu) \cos (2\alpha J)]/2 \;, \nonumber  \\
\tilde N_2 &= 4\sinh (\beta \mu) \cosh (\beta \mu)  \sin (\alpha J)    \cos (\alpha J)
= \sinh (2\beta \mu) \sin (2\alpha J)\;. 
\end{align}
One can notice the first line that is different from the result \eqref{SAGeneralNSLargeL}. This is due to the cosine factor in front of the theta function \eqref{Theta2}. Now one can check that this reduces to the case with current ($\mu=0$) evaluated in \eqref{EELargeSpaceLimitRSector}, not to mention to the case with only chemical potential ($\alpha = J=0$) evaluated in \eqref{SAmuRJLargeL2}. \\

\paragraph{High temperature limit:} 
We compute the entropies \eqref{GeneralEEFormulaJmuHighT} that is valid for the high temperature limit $\beta \to 0$. We set $a+J=0$ for the periodic boundary condition. Then 
\begin{align}
\tilde S_n^{\mu,J}
= \frac{1}{1 \- n} \sum_{k=-\frac{n-1}{2}}^{\frac{n-1}{2}} \log \Big|\frac{e^{-i\frac{\pi}{\tau} (\frac{k \ell_t}{n 2\pi L}+ \tau_1 J + i \tau_2 \mu)^2}}{e^{-i\frac{\pi}{\tau} ( \tau_1 J + i \tau_2 \mu)^2}} \frac{\vartheta_4(\frac{k}{n} \frac{\ell_t}{2\pi L} \frac{1}{\tau} + \frac{\tau_1 J+ i \tau_2 \mu}{\tau}|-\frac{1}{\tau})}{\vartheta_4(\frac{\tau_1 J + i \tau_2 \mu}{\tau}|-\frac{1}{\tau})} \Big|^2  \;.
\end{align} 
Due to the presence of $\alpha$, one can take two different high temperature limits. 

For $\alpha=0$, we have the R\'enyi entropy with the similar computations done is the NSNS case. 
\begin{align}\label{RenyiSAmuRHighT2}
\tilde S_n^{\mu,J} &= \frac{n+1}{3n \beta} \frac{\ell_t^2}{4L^2} -\frac{4}{1 \- n} \sum_{m,l=1}^{\infty} \frac{1}{l}  \frac{\cos [2\pi l\mu ]}{e^{\frac{4\pi^2 l}{\beta} (m-\frac{1}{2})} } \left[\sinh \left( \frac{\pi l}{\beta} \frac{\ell_t}{L} \right) \text{csch} \left( \frac{\pi l}{\beta} \frac{\ell_t}{Ln} \right) -n\right] \;.  
\end{align} 
This reduces to \eqref{SAmuNSHighTR} for $n\to 1$ limit and \eqref{RenyiHighTCurrent12} for $\mu \to 0$, and serves as consistency checks.  

For $\alpha\neq 0$ case. We take $\beta=0$ to get the dominant contribution. The R\'nyi entropy has the form
\begin{align}
\tilde S_n^{\mu,J}
= \frac{1}{1 \- n} \sum_{k=-\frac{n-1}{2}}^{\frac{n-1}{2}} \log \Big|\frac{e^{-i\frac{2\pi^2}{\alpha} (\frac{k \ell_t}{n 2\pi L}+ \frac{\alpha}{2\pi} J )^2}}{e^{-i\frac{2\pi^2}{\alpha} ( \frac{\alpha}{2\pi} J )^2}} \frac{\vartheta_4(\frac{k}{n} \frac{\ell_t}{L} \frac{1}{\alpha} + J|-\frac{2\pi}{\alpha})}{\vartheta_4(J|-\frac{2\pi}{\alpha})} \Big|^2  \;.
\end{align} 
Similar to the NSNS sector, the first factor inside the log is pure imaginary and would not contribute.  Due to the presence of $\alpha$, one can take the zero temperature limit in a straightforward manner as $ \frac{1}{\tau}  \to \frac{2\pi}{\alpha} $. 
The R\'enyi entropy reads
\begin{align}\label{RenyiSAmuRHighT1}
\tilde S_n^{\mu,J} =
 \frac{2}{1 \- n} \sum_{m=1}^{\infty} \sum_{k=-\frac{n-1}{2}}^{\frac{n-1}{2}} \log \Big[\frac{\cos (2\pi J) - \cos (\frac{2\pi}{\alpha} [m-\frac{1}{2}] - \frac{1}{\alpha} \frac{k}{n} \frac{\ell_t}{L}) }{\cos (2\pi J) - \cos (\frac{2\pi}{\alpha} [m-\frac{1}{2}])} \Big]
\;. 
\end{align} 
The result reduces to \eqref{SAmuNSHighTRAlphaZero1} and \eqref{RenyiSAmuRJHighT}, 
and is similar to that of the anti-periodic fermion. Only difference is the relative sign between two cosine terms in the log. While the sum over $k$ is not easy to evaluate, it is clear that the individual term contributes with the periodic behavior. There are further sub-leading contributions as $\beta \to 0$. It will be interesting to perform a numerical study for this limit. \\

\subsection{With only modulus parameters}\label{sec:EEModulus}

For completeness, we present the results without the chemical potential and current $\mu=J=0$, yet keeping the modulus parameters $\tau = \tau_1 + i\tau_2 = \frac{1}{2\pi}(\alpha+i \beta)\neq 0$. Here we only consider the entanglement entropy, while it is straightforward to generalize to the R\'enyi entropy.  

\subsubsection{Anti-periodic fermion}

Let us consider the anti-periodic fermion first, $a=1/2$. From \eqref{GeneralEEFormulaJmu}, one can get the entanglement entropy with current dependence  
\begin{align}
S &= \frac{1}{1\- n} \sum_{k=-\frac{n-1}{2}}^{\frac{n-1}{2}} \log \Big|\frac{\vartheta_3 (\frac{k}{n} \frac{\ell_t}{2\pi L} |\tau )}{\vartheta_3 (0 |\tau)} \Big|^2  \Bigg]_{n=1}\;. 
\end{align} 

For the low temperature limit, $\beta \to \infty$, the entanglement entropy reads  
\begin{align}
S_A 
&= 4 \sum_{l,m=1}^{\infty} \!\!\! \frac{(-1)^{l-1}}{l} \frac{ \cos (\alpha l [m-1/2])}{e^{ (m-1/2) \beta l}}  \left[1\- l \frac{\ell_t}{2L} \cot \left(\!l \frac{\ell_t}{2L} \right) \right]. 
\end{align} 
We note that the entanglement entropy has separate contributions from both the modulus parameters  $\alpha$ and $\beta$.

For the large radius limit $\ell_t/L \ll 1$, we also compute the entanglement entropy as 
\begin{align}\label{SAGeneralNSLargeLNoGaugeFields}
S &= \frac{2}{3} \frac{\ell_t^2}{4L^2} \sum_{m=1}^\infty  
\frac{\cosh ([m\-\frac{1}{2}]\beta) \cos ([m\-\frac{1}{2}]\alpha) + 1 }{[\cosh ([m\-\frac{1}{2}]\beta) \cos ([m\-\frac{1}{2}]\alpha) \+ 1]^2 \+ [\sinh ([m\-\frac{1}{2}]\beta) \sin ([m\-\frac{1}{2}]\alpha)]^2} \nonumber \\
&  \+ \mathcal O(\frac{\ell_t^2}{L^2})^4 \;. 
\end{align} 
Thus it still vanishes as $\mathcal O\left(\frac{\ell_t^2}{L^2}\right)$ in the large radius limit. This tells that the entanglement entropy is actually independent of the spin structures or twisted boundary conditions at infinite space limit. This is also true for the periodic boundary condition. It will be interesting to explorer this for more general twisted boundary condition when $ 0<a<1/2$. 

For the high temperature limit, the R\'enyi entropy for $\alpha=0$ is given by 
\begin{align}\label{RenyiSAmuNSJHighTOnlyTau}
\tilde S_n
&\= \frac{(n\+1)}{3n} \frac{1}{\beta} \frac{\ell_t^2}{4L^2} + \sum_{l=1}^{\infty} \frac{(-1)^{l-1}}{l \sinh \left( \frac{4\pi^2 }{\beta}l\right)} 
   \frac{2}{1\- n}\left[ \sinh \left(\!{\frac{\pi l}{\beta} \frac{ \ell_t}{L}}\!\right) \text{csch} \left(\!{\frac{\pi l}{\beta} \frac{\ell_t}{nL}}\!\right) \- n\right]  \;.
\end{align} 
This result is consistent with \eqref{RenyiSAmuNSJHighTGeneral}. Entanglement entropy is straightforward to evaluate. 
For $\alpha \neq 0$, we can get the dominant contribution by taking $\beta\to 0$. Thus the R\'enyi entropy reads
\begin{align}\label{RenyiSAmuNSJHighOnlyTau}
\tilde S_n^J =
 \frac{2}{1 \- n} \sum_{m=1}^{\infty} \sum_{k=-\frac{n-1}{2}}^{\frac{n-1}{2}} \log \Big[\frac{\cos (2\pi J) + \cos (\frac{2\pi}{\alpha} [m-\frac{1}{2}] - \frac{1}{\alpha} \frac{k}{n} \frac{\ell_t}{L}) }{\cos (2\pi J) + \cos (\frac{2\pi}{\alpha} [m-\frac{1}{2}])} \Big]
\;. 
\end{align} 
This result is consistent with \eqref{RenyiSAmuNSJHighT00}.

\subsubsection{Periodic fermion}

Let us consider the periodic fermion, $a=0$. From \eqref{GeneralEEFormulaJmu}, one can get the entanglement entropy with the chemical potential and current dependence  
\begin{align}
S &= \frac{1}{1\- n} \sum_{k=-\frac{n-1}{2}}^{\frac{n-1}{2}} \log \Big|\frac{\vartheta_2 (\frac{k}{n} \frac{\ell_t}{2\pi L} |\tau )}{\vartheta_2 (0 |\tau)} \Big|^2  \Bigg]_{n=1}\;. 
\end{align} 

For the low temperature limit, $\beta \to \infty$, the entanglement entropy is given by 
\begin{align}
S
&= 2 \sum_{l=1}^{\infty} \!\! \frac{(-1)^{l-1}}{l}  \left[ 1+ 2 \sum_{m=1}^{\infty} \frac{ \cos (\alpha l m)}{e^{ m \beta l}} \right] \left[1\- l \frac{\ell_t}{2L} \cot \left(\! l \frac{\ell_t}{2L} \right) \right]. 
\end{align} 
Thus it has non-vanishing contribution at zero temperature. 

For the large radius limit $\ell_t/L \ll 1$, the entanglement entropy is 
\begin{align}\label{SAGeneralRLargeL}
S
= \frac{2}{3} \frac{\ell_t^2}{4L^2} & \Bigg(\frac{1}{4} +  
\sum_{m=1}^\infty  \left[ 
\frac{\cosh (m \beta) \cos (m\alpha) + 1 }{[\cosh (m \beta) \cos (m\alpha) + 1]^2 + [\sinh (m\beta) \sin (m\alpha)]^2} \right] \Bigg) + \mathcal O(\frac{\ell_t^2}{L^2})^4 \;. 
\end{align} 
Thus it still vanishes as $\frac{\ell_t^2}{L^2}$ in the large radius limit. 

We compute the entropies \eqref{GeneralEEFormulaJmuHighT} that is valid for the high temperature limit $\beta \to 0$. We set $a=J=\mu=0$ for the periodic boundary condition without background fields. Then 
\begin{align}\label{RenyiSAmuRHighT20nlyTauGeneral}
\tilde S_n
= \frac{1}{1 \- n} \sum_{k=-\frac{n-1}{2}}^{\frac{n-1}{2}} \log \Big|e^{-i\frac{\pi}{\tau} (\frac{k \ell_t}{n 2\pi L})^2} \frac{\vartheta_4(\frac{k}{n} \frac{\ell_t}{2\pi L} \frac{1}{\tau} |-\frac{1}{\tau})}{\vartheta_4(0|-\frac{1}{\tau})} \Big|^2  \;.
\end{align} 
Due to the presence of $\alpha$, one can take two different high temperature limits. 
For $\alpha=0$, we have the R\'enyi entropy with the similar computations done is the NSNS case. 
\begin{align}\label{RenyiSAmuRHighT20nlyTau}
\tilde S_n &= \frac{1}{1 \- n} \sum_{k=-\frac{n-1}{2}}^{\frac{n-1}{2}} \log \Big|e^{-\frac{2\pi^2}{\beta} (\frac{k \ell_t}{n 2\pi L})^2}  \frac{\vartheta_4(-i\frac{k}{n} \frac{\ell_t}{L} \frac{1}{\beta} |i\frac{2\pi}{\beta})}{\vartheta_4(0|i\frac{2\pi}{\beta})} \Big|^2 \nonumber\\
&= \frac{n+1}{3n \beta} \frac{\ell_t^2}{4L^2} -\frac{2}{1 \- n} \sum_{l=1}^{\infty} \frac{1}{l}  \frac{1}{\sinh \left(\frac{2\pi^2 l}{\beta}\right)}  \left[\sinh \left( \frac{\pi l}{\beta} \frac{\ell_t}{L} \right) \text{csch} \left( \frac{\pi l}{\beta} \frac{\ell_t}{Ln} \right) -n\right] \;.  
\end{align} 
This is consistent with \eqref{RenyiSAmuRHighT2}.

Now we turn to $\alpha\neq 0$ case. To get the dominant contribution, we take $\beta=0$. The R\'nyi entropy \eqref{RenyiSAmuRHighT20nlyTauGeneral} has the form
\begin{align}\label{RenyiSAmuRHighT1OnlyTau}
\tilde S_n 
&= \frac{2}{1 \- n} \sum_{m=1}^{\infty} \sum_{k=-\frac{n-1}{2}}^{\frac{n-1}{2}} \log \Big[\frac{1 - \cos (\frac{2\pi}{\alpha} [m-\frac{1}{2}] - \frac{1}{\alpha} \frac{k}{n} \frac{\ell_t}{L}) }{1 - \cos (\frac{2\pi}{\alpha} [m-\frac{1}{2}])} \Big]
\;. 
\end{align} 
The result is consistent with \eqref{RenyiSAmuRHighT1} by setting the background fields to vanish $\mu=J=0$. While the sum over $k$ is not easy to evaluate, it is clear that the individual term contributes with the periodic behavior. There are further sub-leading contributions as $\beta \to 0$. It will be interesting to perform a numerical study for this limit. \\

\section{Mutual information}\label{sec:MutualInformation} 

Mutual (R\'enyi) information measures the entanglement between two intervals, $A$ and $B$, of length $\ell_A$ and $\ell_B$ separated by $\ell_C$. It is given by 
\begin{align}\label{defMutualInfomation}
I_n(A,B) = S_n (A) + S_n (B) - S_n (A\cup B) \;.
\end{align}
This is a finite quantity, free of UV divergences. Mutual (R\'enyi) information turns out to share the same functional dependences on the current $J$ and the chemical potential $\mu$ as those of the R\'enyi and the entanglement entropies. 

Using \eqref{defMutualInfomation} and the results given in \S \ref{sec:EEMuJ}, we can obtain the general formula for the mutual R\'enyi information in the presence of the background fields. 
Similar to the entropies, the mutual information factories into two different contributions. 
\begin{align}
I_n (A,B) &= I^{0}_n (A,B) + I^{\mu,J}_n (A,B) \;.
\end{align} 
The first contribution is independent of the background fields \cite{Herzog:2013py}. 
\begin{align}\label{GeneralMutualInformation0}
I^{0}_n (A,B) &=\frac{1}{1\- n} \sum_{k=-\frac{n-1}{2}}^{\frac{n-1}{2}} \log \bigg. \Big| \frac{\vartheta [\substack{1/2 \\ 1/2 }](\frac{\ell_A+\ell_B+\ell_C}{2\pi L}|\tau) ~\vartheta [\substack{1/2 \\ 1/2 }](\frac{\ell_C}{2\pi L}|\tau)}{\vartheta [\substack{1/2 \\ 1/2 }](\frac{\ell_A+\ell_C}{2\pi L}|\tau) ~\vartheta [\substack{1/2 \\ 1/2 }](\frac{\ell_B+\ell_C}{2\pi L}|\tau)} \Big|^{2\frac{k^2}{n^2}} \nonumber \\
&=-\frac{n+1}{6n} \log \bigg. \Big| \frac{\vartheta [\substack{1/2 \\ 1/2 }](\frac{\ell_A+\ell_B+\ell_C}{2\pi L}|\tau) ~\vartheta [\substack{1/2 \\ 1/2 }](\frac{\ell_C}{2\pi L}|\tau)}{\vartheta [\substack{1/2 \\ 1/2 }](\frac{\ell_A+\ell_C}{2\pi L}|\tau) ~\vartheta [\substack{1/2 \\ 1/2 }](\frac{\ell_B+\ell_C}{2\pi L}|\tau)} \Big|\;. 
\end{align} 
This has been studied, and we focus on the other contribution. 
The second one has all the dependences on the current and chemical potential. 
\begin{align}\label{GeneralMutualInformation}
I^{\mu,J}_n 
&=\frac{1}{1\- n} \sum_{k=-\frac{n-1}{2}}^{\frac{n-1}{2}} \!\!\!\! \log  \Big| \frac{\vartheta [\substack{1/2-a-J \\ b-1/2 }](\frac{k}{n} \frac{\ell_A}{2\pi L} \+ \tau_1 J \+ i \tau_2 \mu|\tau)~ \vartheta [\substack{1/2-a-J \\ b-1/2 }](\frac{k}{n} \frac{\ell_B}{2\pi L} \+ \tau_1 J \+ i \tau_2 \mu|\tau)}{\vartheta [\substack{1/2-a-J \\ b-1/2 }](\tau_1 J \+ i \tau_2 \mu|\tau)~\vartheta [\substack{1/2-a-J \\ b-1/2 }](\frac{k}{n} \frac{\ell_A+\ell_B}{2\pi L} \+ \tau_1 J \+ i \tau_2 \mu|\tau)} \Big|^2 \;. 
\end{align} 
It is interesting to note that the parts of the mutual information that depend on the current and the chemical potential are actually independent of $\ell_C$, the separation distance between the two sub-systems. This is even more clearer when we evaluate the information explicitly below. We note that it would be interesting to find out some special cases where $I^{\mu,J}_n$ dominates over $I^{0}_n$, so that the chemical potential and current dependences would be clearly visible. 
 
We provide detailed studies of it for the Dirac fermion as a function of chemical potential or/and current on a torus in the low temperature, large radius, and high temperature limits, in turn. For each limit, the computation is similar to the previous cases, and we focus on the results and their physical properties. We organize each sub-section by presenting the results for NSNS sector, followed by RNS sector.

\subsection{Low temperature limit} 

In this section, we consider the mutual information of the Dirac fermion in the low temperature limit, $\beta \to \infty$. We evaluate the NSNS sector and RNS sector in turn.  

\paragraph{Anti-periodic fermion:}
In the low temperature limit, we compute the Mutual R\'enyi information for the NSNS sector for both the anti-periodic boundary conditions on spatial and temporal circles. We use the formula \eqref{GeneralMutualInformation} with $b=1/2, a+J=1/2$ and take the low temperature limit $\beta \to \infty$. 
\begin{align}\label{MutualInformationLowTNS}
I^{\mu,J}_n (A,B) 
&\=\frac{1}{1\- n} \sum_{k=-\frac{n-1}{2}}^{\frac{n-1}{2}} \log  \Big| \frac{\vartheta _3(\frac{k}{n} \frac{\ell_A}{2\pi L} + \tau_1 J + i \tau_2 \mu|\tau)~ \vartheta_3(\frac{k}{n} \frac{\ell_B}{2\pi L} + \tau_1 J + i \tau_2 \mu|\tau)}{\vartheta_3(\tau_1 J + i \tau_2 \mu|\tau)~\vartheta_3(\frac{k}{n} \frac{\ell_A+\ell_B}{2\pi L} + \tau_1 J + i \tau_2 \mu|\tau)} \Big|^2 \nonumber \\
&\=\frac{2}{n\-1}\! \sum_{m,l=1}^{\infty} \frac{(-1)^{l\-1}}{l e^{\beta l (m-1/2)}} 
 \! \left[e^{-\beta\mu l} \cos (\alpha l [m-\frac{1}{2}] \+ \alpha J l) \+ e^{\beta\mu l} \cos (\alpha l [m-\frac{1}{2}] \- \alpha J l)\right] \nonumber \\
&\qquad\qquad \times  \bigg[n-  \frac{\sin \left(\! \frac{l\ell_A}{2L} \!\right)}{\sin \left(\! \frac{l}{n} \frac{\ell_A}{2L} \!\right)} -  \frac{\sin \left(\! \frac{l\ell_B}{2L} \!\right)}{\sin \left(\! \frac{l}{n} \frac{\ell_B}{2L} \!\right)} +  \frac{\sin \big(\! \frac{l(\ell_A+\ell_B)}{2L} \big)}{\sin \left(\! \frac{l}{n} \frac{\ell_A+\ell_B}{2L} \!\right)} \bigg] \;. 
\end{align} 
This result is valid for $-1/2 < \mu < 1/2$. For other values of $\mu$, one can get modified results similar to those of the entropies. The computation is straightforward.  
The corresponding mutual information can be evaluated by taking $n \to 1$ limit. 
\begin{align}
I^{\mu,J} (A,B) 
&\= 2\! \sum_{m,l=1}^{\infty} \frac{(-1)^{l\-1}}{l e^{\beta l (m-1/2)}} 
 \! \left[e^{-\beta\mu l} \cos (\alpha l [m-\frac{1}{2}] \+ \alpha J l) \+ e^{\beta\mu l} \cos (\alpha l [m-\frac{1}{2}] \- \alpha J l)\right] \nonumber \\
&\qquad\times \! \bigg[1 \- \frac{l}{2L} \left\{ \ell_A \cot \left(\! \frac{l\ell_A}{2L} \! \right) \+ \ell_B \cot \left(\! \frac{l\ell_B}{2L} \! \right) \- (\ell_A \+\ell_B)  \cot \left(\! \frac{l(\ell_A \+\ell_B)}{2L} \!\right) \! \right\} \!\bigg] \;. 
\end{align} 
There are several interesting observations to make here. First, we confirm that the mutual (R\'enyi) information $I^{\mu,J}_n (A,B)$ is independent of $\ell_C$, and thus the separation between the two sub-systems, while the part $I^{0} (A,B)$ depends on $\ell_C$. Second, $I^{\mu,J}_n (A,B)$ has the same functional dependences on $\mu$ and $J$ as the entropies. This can be explicitly checked with the results given in \eqref{RenyiEELowTAntiNS} and \eqref{EELowTAntiNS}. Third, we compute the mutual information for the special case $\ell_A = \ell_B = \tilde \ell_t $. 
\begin{align}
I^{\mu J} 
&= 2\!\! \sum_{l,m=1}^{\infty} \!\!\! \frac{(-1)^{l-1}}{l} \frac{ \cos (\alpha l [J\!\-m\+\frac{1}{2}])}{e^{ (m-\frac{1}{2}-\mu) \beta l}} \! \left[ 1  \+ e^{-2l\beta \mu} \frac{\cos (\alpha l[J\+m\-\frac{1}{2}])}{\cos (\alpha l [J\-m\+\frac{1}{2}])} \right]\!\! \left[1\- l \frac{\tilde \ell_t}{L} \csc \left(\! l \frac{\tilde \ell_t}{L} \!\right) \right] \;. 
\end{align} 
This can be compared with \eqref{EELowTAntiNS}. The mutual information is identical to the entanglement entropy, except the dependence on the subsystem sizes.

For the special case $J=\alpha=0$, we get the chemical potential dependent part of the Mutual R\'enyi information.
\begin{align}
I^\mu_n &
=\frac{2}{n\-1}\! \sum_{l=1}^{\infty} \frac{(-1)^{l\-1}}{l} 
\frac{\cosh (l\beta\mu)}{\sinh ( \frac{l\beta}{2} )}  \bigg[n\-  \frac{\sin \left(\! \frac{l\ell_A}{2L}\!\right)}{\sin \left(\! \frac{l}{n} \frac{\ell_A}{2L}\!\right)} \-  \frac{\sin \left(\! \frac{l\ell_B}{2L}\!\right)}{\sin \left(\! \frac{l}{n} \frac{\ell_B}{2L}\!\right)} +  \frac{\sin \left(\! \frac{l(\ell_A+\ell_B)}{2L}\!\right)}{\sin \left(\! \frac{l}{n} \frac{\ell_A+\ell_B}{2L}\!\right)} \bigg] \;. 
\end{align} 
The result is only valid for $-1/2 < \mu < 1/2$. For the other values of $\mu$, it is straightforward to evaluate following the previous discussion. 
On the other hand, the current dependent Mutual R\'enyi information can be obtained by $\mu=0$. 
\begin{align}
I^{J}_n 
&\=\frac{4}{n\-1}\! \sum_{m,l=1}^{\infty} \frac{(-1)^{l\-1}}{l e^{\beta l (m-1/2)}} 
~ \cos (\alpha l [m-\frac{1}{2}]) \cos (\alpha J l)  \bigg[n- \cdots  \bigg] \;,  
\end{align} 
where the $\cdots$ represent the same dependence of sine function given in the Mutual R\'enyi information in \eqref{MutualInformationLowTNS}. We find that the mutual (R\'enyi) information has the same dependence on $\mu$ and $J$ as the entropies.

\paragraph{Periodic fermion:}
The Mutual R\'enyi information for the RNS sector for the periodic boundary condition on the spatial circle can be computed in a similar way. We use $b=1/2, a+J=0 $
\begin{align}\label{MutualInformationLowTR}
I^{\mu,J}_n (A,B) 
&\=\frac{1}{1\- n} \sum_{k=-\frac{n-1}{2}}^{\frac{n-1}{2}} \log  \Big| \frac{\vartheta _2(\frac{k}{n} \frac{\ell_A}{2\pi L} + \tau_1 J + i \tau_2 \mu|\tau)~ \vartheta_2(\frac{k}{n} \frac{\ell_B}{2\pi L} + \tau_1 J + i \tau_2 \mu|\tau)}{\vartheta_2(\tau_1 J + i \tau_2 \mu|\tau)~\vartheta_2(\frac{k}{n} \frac{\ell_A+\ell_B}{2\pi L} + \tau_1 J + i \tau_2 \mu|\tau)} \Big|^2 \nonumber \\
&\=\frac{2}{n\-1}\! \sum_{l=1}^{\infty} \frac{(-1)^{l\-1}}{l } ~\bigg[n-  \frac{\sin \left(\! \frac{l\ell_A}{2L} \!\right)}{\sin \left(\! \frac{l}{n} \frac{\ell_A}{2L} \!\right)} -  \frac{\sin \left(\! \frac{l\ell_B}{2L} \!\right)}{\sin \left(\! \frac{l}{n} \frac{\ell_B}{2L} \!\right)} +  \frac{\sin \big(\! \frac{l(\ell_A+\ell_B)}{2L} \big)}{\sin \left(\! \frac{l}{n} \frac{\ell_A+\ell_B}{2L} \!\right)} \bigg]
 \nonumber \\
&\times  \Big[ e^{-\beta\mu l} \cos ( \alpha J l)  + \sum_{m=1}^{\infty} e^{-\beta l m}  \big\{ e^{-\beta\mu l} \cos (\alpha l [m+J]) \+ e^{\beta\mu l} \cos (\alpha l [m-J]) \big\} \Big] 
\;. 
\end{align} 
This result is valid for $0 \leq \mu < 1$. The corresponding mutual information can be evaluated by taking $n \to 1$ limit. 
\begin{align}
I^{\mu,J} (A,B) 
&\= 2\! \sum_{l=1}^{\infty} \frac{(-1)^{l\-1}}{l } 
 \! \Big[ \frac{\cos ( \alpha J l)}{e^{\beta\mu l}}  + \sum_{m=1}^{\infty} e^{-\beta l m}  \big\{ e^{-\beta\mu l} \cos (\alpha l [m+J]) \+ e^{\beta\mu l} \cos (\alpha l [m-J]) \big\} \Big] \nonumber \\
&\qquad\times \! \bigg[1 \- \frac{l}{2L} \left\{ \ell_A \cot \left(\! \frac{l\ell_A}{2L} \! \right) \+ \ell_B \cot \left(\! \frac{l\ell_B}{2L} \! \right) \- (\ell_A \+\ell_B)  \cot \left(\! \frac{l(\ell_A \+\ell_B)}{2L} \!\right) \! \right\} \!\bigg] \;. 
\end{align} 
The mutual (R\'enyi) information $I^{\mu,J}_n (A,B)$ is independent of $\ell_C$ and has the same functional dependences on $\mu$ and $J$ as the entropies, similar to the anti-periodic case. 

Compared to the mutual (R\'enyi) information of the NSNS sector given in \eqref{MutualInformationLowTNS}, that of the RNS sector has a distinct contribution that is proportional to $ \cos ( \alpha J l)$ in the first line. Interestingly, this term provide a non-zero contribution in the zero temperature limit, $\beta\mu =\tilde M = const. $ for $\beta \to \infty, \mu \to 0 $. 
\begin{align}
2 \sum_{l=1}^{\infty} \frac{(-1)^{l\-1}}{l } 
\frac{\cos ( \alpha J l)}{e^{\tilde M l}} \bigg[1 \- \frac{l}{2L} \left\{ \ell_A \cot \left(\! \frac{l\ell_A}{2L} \! \right) \+ \ell_B \cot \left(\! \frac{l\ell_B}{2L} \! \right) \- (\ell_A \+\ell_B)  \cot \left(\! \frac{l(\ell_A \+\ell_B)}{2L} \!\right) \! \right\} \!\bigg] \;. 
\end{align} 
The mutual (R\'enyi) information turns out to be non-zero for 
\begin{align}
\mu = \frac{N}{2} \;, 
\end{align}
which can be identified as the energy levels of the Dirac fermion on a circle. This is explained in detail in the previous section \S \ref{sec:EEMu} with $\mu\neq 0$ and $J=0$. 

For the special case $J=\alpha=0$, we get the chemical potential dependent part of the Mutual R\'enyi information.
\begin{align}
I^\mu_n &
=\frac{2}{n\-1}\! \sum_{l=1}^{\infty} \frac{(-1)^{l\-1}}{l} \! \left(\! e^{-\beta \mu l} \+ 
\frac{\cosh (l\beta\mu)}{e^{\frac{\beta l}{2}}\sinh ( \frac{l\beta}{2} )} \!\right) \! 
\bigg[n\-  \frac{\sin \left(\! \frac{l\ell_A}{2L}\!\right)}{\sin \left(\! \frac{l}{n} \frac{\ell_A}{2L}\!\right)} \-  \frac{\sin \left(\! \frac{l\ell_B}{2L}\!\right)}{\sin \left(\! \frac{l}{n} \frac{\ell_B}{2L}\!\right)} +  \frac{\sin \left(\! \frac{l(\ell_A+\ell_B)}{2L}\!\right)}{\sin \left(\! \frac{l}{n} \frac{\ell_A+\ell_B}{2L}\!\right)} \bigg] \;. 
\end{align} 
The result is only valid for $0 \leq \mu < 1$. For the other values of $\mu$, it is straightforward to evaluate following the previous discussion. 
On the other hand, the current dependent Mutual R\'enyi information can be obtained by $\mu=0$. 
\begin{align}
I^{J}_n 
&\=\frac{4}{n\-1}\! \sum_{l=1}^{\infty} \frac{(-1)^{l\-1}}{l } \left( \cos(\alpha J l) + 
2 \sum_{m=1}^{\infty} e^{-\beta l m} \cos (\alpha l m) \cos (\alpha J l) \right) 
\bigg[n- \cdots  \bigg] \;,  
\end{align} 
where the $\cdots$ represent the same dependence of sine function given in the Mutual R\'enyi information in \eqref{MutualInformationLowTNS}. We find that the mutual R\'enyi information has the same dependence on $\mu$ and $J$. The first term in the bracket is special for the RNS sector.

\subsection{Large radius limit}
 
The large radius limit of the mutual (R\'enyi) information is straightforward to evaluate. The computation becomes much more complicated compared to the entropies. It is relatively easy to see that the Mutual information also vanishes at least $\mathcal O \left( (\frac{\ell_A}{L})^2,(\frac{\ell_B}{L})^2, \frac{\ell_A \ell_B}{L^2} \right)$. We are not going to explicitly write the result here.

\subsection{High temperature limit} 

For the analysis in the high temperature limit, the general formulas \eqref{GeneralMutualInformation0} and \eqref{GeneralMutualInformation} can be rewritten by using the modular transformation. The mutual information \eqref{GeneralMutualInformation0} that is independent of the background gauge fields slightly change 
\begin{align}\label{GeneralMutualInformation0HighT}
\tilde I^{0}_n (A,B) &=-\frac{n+1}{6n} \log \bigg. \Big| \frac{e^{-\frac{i\pi}{\tau}(\frac{\ell_A+\ell_B+\ell_C}{2\pi L} )^2-\frac{i\pi}{\tau}(\frac{\ell_C}{2\pi L})^2}}{e^{-\frac{i\pi}{\tau}(\frac{\ell_A+\ell_C}{2\pi L})^2 -\frac{i\pi}{\tau}(\frac{\ell_B+\ell_C}{2\pi L})^2}}
\frac{\vartheta [\substack{1/2 \\ 1/2 }](\frac{\ell_A+\ell_B+\ell_C}{2\pi L \tau}|\frac{-1}{\tau}) ~\vartheta [\substack{1/2 \\ 1/2 }](\frac{\ell_C}{2\pi L \tau}|\frac{-1}{\tau})}{\vartheta [\substack{1/2 \\ 1/2 }](\frac{\ell_A+\ell_C}{2\pi L \tau}|\frac{-1}{\tau}) ~\vartheta [\substack{1/2 \\ 1/2 }](\frac{\ell_B+\ell_C}{2\pi L \tau}|\frac{-1}{\tau})} \Big| \;. 
\end{align}  

The high temperature limit is sensitive to the parameter $\alpha=2\pi \tau_1$. For $\alpha \neq 0$, it is straightforward to evaluate the zero temperature limit $\beta \to 0$ by using $ \frac{1}{\tau} = \frac{2\pi}{\alpha + i \beta}  \to \frac{2\pi}{\alpha} $. We get 
\begin{align}\label{MutualInfoNoBGFHighTAlphaNonZero}
\tilde I^{0}_n &=-\frac{n+1}{6n} \log \bigg[ \frac{\sin (\pi \frac{\ell_A+\ell_B+\ell_C}{L \alpha} ) \sin (\pi \frac{\ell_C}{L \alpha})}{\sin (\pi \frac{\ell_A+\ell_C}{L \alpha} ) \sin (\pi \frac{\ell_B+\ell_C}{L \alpha})} \nonumber \\
& \qquad\qquad\qquad \times \prod_{m=1}^\infty
\frac{\big[ \cos (2\pi \frac{\ell_A+\ell_B+\ell_C}{L \alpha} ) \- \cos (\frac{4\pi^2}{\alpha} m)\big]
\big[ \cos (2\pi \frac{\ell_C}{L \alpha} ) \- \cos (\frac{4\pi^2}{\alpha} m)\big]}{\big[ \cos (2\pi \frac{\ell_A+\ell_C}{L \alpha} m) \- \cos (\frac{4\pi^2}{\alpha} )\big] 
\big[ \cos (2\pi \frac{\ell_B+\ell_C}{L \alpha} ) - \cos (\frac{4\pi^2}{\alpha} m )\big]} \bigg] \;. 
\end{align} 
For $\alpha=0$, the oscillating behavior changes into a decaying one. Thus we get 
\begin{align}\label{MutualInfoNoBGFHighTAlphaZero}
\tilde I^{0}_n &=\frac{n+1}{6n} \frac{1}{\beta} \frac{\ell_A \ell_B}{L^2}
-\frac{n+1}{6n} \log \bigg[ \frac{\sinh (\pi \frac{\ell_A+\ell_B+\ell_C}{L \beta} ) \sinh (\pi \frac{\ell_C}{L\beta})}{\sinh (\pi \frac{\ell_A+\ell_C}{L \beta} ) \sinh (\pi \frac{\ell_B+\ell_C}{L\beta})} \nonumber \\
& \qquad\qquad\qquad \times \prod_{m=1}^\infty
\frac{\big[ \cosh (2\pi \frac{\ell_A+\ell_B+\ell_C}{L \beta} ) \- \cosh (\frac{4\pi^2}{\beta}m )\big]
\big[ \cosh (2\pi \frac{\ell_C}{L \beta} ) \- \cosh (\frac{4\pi^2}{\beta} m )\big]}{\big[ \cosh (2\pi \frac{\ell_A+\ell_C}{L \beta} ) \- \cosh (\frac{4\pi^2}{\beta} m)\big] 
\big[ \cosh (2\pi \frac{\ell_B+\ell_C}{L \beta} ) - \cos (\frac{4\pi^2}{\beta} m)\big]} \bigg] \;. 
\end{align} 

The chemical potential and current dependent mutual information has further contributions as 
\begin{align}\label{GeneralMutualInformationHighT}
\tilde I^{\mu,J}_n 
&=\frac{1}{1\- n} \sum_{k=-\frac{n-1}{2}}^{\frac{n-1}{2}} \log  \Big| \frac{e^{-\frac{i\pi}{\tau}(\frac{k}{n} \frac{\ell_A}{2\pi L} \+ \tau_1 J + i \tau_2 \mu)^2}~ e^{-\frac{i\pi}{\tau}(\frac{k}{n} \frac{\ell_B}{2\pi L} \+ \tau_1 J + i \tau_2 \mu)^2}}{e^{-\frac{i\pi}{\tau}(\tau_1 J + i \tau_2 \mu)^2}~ e^{-\frac{i\pi}{\tau}(\frac{k}{n} \frac{\ell_A+\ell_B}{2\pi L} \+ \tau_1 J + i \tau_2 \mu)^2}} \nonumber \\
&\qquad\qquad \times \frac{\vartheta [\substack{1/2-b \\ 1/2-a-J }](\frac{k}{n} \frac{\ell_A}{2\pi L \tau} \+ \frac{\tau_1 J + i \tau_2 \mu}{\tau}|\frac{-1}{\tau})~ \vartheta [\substack{1/2-b \\ 1/2-a-J  }](\frac{k}{n} \frac{\ell_B}{2\pi L \tau} \+ \frac{\tau_1 J + i \tau_2 \mu}{\tau}|\frac{-1}{\tau})}{\vartheta [\substack{1/2-b \\ 1/2-a-J  }](\frac{\tau_1 J + i \tau_2 \mu}{\tau}|\frac{-1}{\tau})~\vartheta [\substack{1/2-b \\ 1/2-a-J  }](\frac{k}{n} \frac{\ell_A+\ell_B}{2\pi L \tau} \+ \frac{\tau_1 J + i \tau_2 \mu}{\tau}|\frac{-1}{\tau})} \Big|^2 \;. 
\end{align} 
We study this mutual information in some details in this section.

\paragraph{Anti-periodic fermion:}
For the anti-periodic fermion in the NSNS sector, we have $ a+J = 1/2, b=1/2$. 
For $\alpha=0$, we have the following result for the mutual R\'enyi information 
\begin{align}\label{GeneralMutualInformationHighTAntiAplhaZero11}
I^{\mu,J}_n
&\= -\frac{n+1}{12 n} \frac{1}{\beta} \frac{2\ell_A \ell_B}{L^2} + 4 \sum_{m,l=1}^{\infty} \frac{(-1)^{l\-1}}{l } e^{-\frac{4\pi^2 l}{\beta} (m-\frac{1}{2})} \cos [2\pi\mu l] \nonumber \\
& \times  \frac{1}{n\-1} \bigg[n-  \frac{\sinh \left(\! \frac{\pi}{\beta} \frac{l\ell_A}{L} \!\right)}{\sinh \left(\! \frac{\pi}{\beta} \frac{l}{n} \frac{\ell_A}{L} \!\right)} -  \frac{\sinh \left(\! \frac{\pi}{\beta}\frac{l\ell_B}{L} \!\right)}{\sinh \left(\! \frac{\pi}{\beta} \frac{l}{n} \frac{\ell_B}{L} \!\right)} +  \frac{\sinh \big(\! \frac{\pi}{\beta} \frac{l(\ell_A+\ell_B)}{L} \big)}{\sinh \left(\! \frac{\pi}{\beta} \frac{l}{n} \frac{\ell_A+\ell_B}{L} \!\right)} \bigg] \;. 
\end{align} 
The mutual R\'enyi information for $\alpha=0$ decays except the first term, which can be combined with that of \eqref{MutualInfoNoBGFHighTAlphaZero}. 
The corresponding mutual information is straightforward to compute. 
For $\alpha \neq 0$, we get different results.   
\begin{align}\label{GeneralMutualInformationHighTAntiAplhaZero}
\tilde I^{\mu,J}_n 
&\=\frac{2}{1\- n} \sum_{m=1}^{\infty} \sum_{k=\frac{1-n}{2}}^{\frac{n-1}{2}} \!\! \log \! \Bigg[\! \frac{\big[\! \cos (2\pi [J \+ \frac{k}{n \alpha}\frac{\ell_A}{L}] ) \+ \cos (\frac{4\pi^2}{\alpha} m)\big]\!
\big[\! \cos (2\pi [J \+ \frac{k}{n \alpha}\frac{\ell_B}{L}] ) \+ \cos (\frac{4\pi^2}{\alpha} m))\big]}{\big[\cos (2\pi J ) \+ \cos (\frac{4\pi^2}{\alpha} m)\big] 
\big[ \cos (2\pi [J + \frac{k}{n \alpha}\frac{\ell_A+\ell_B}{L}] ) \+ \cos (\frac{4\pi^2}{\alpha} m)\big]} \Bigg] \;. 
\end{align}

\paragraph{Periodic fermion:}
For the periodic fermion in the RNS sector, we have $ a+J = 0, b=1/2$. The exact expression is similar to the NSNS sector. The only difference comes from the sign change between the $\vartheta_3$ and $\vartheta_4$. Thus for $\alpha=0$, we have the following result 
\begin{align}\label{GeneralMutualInformationHighTPeriAplhaZero11}
I^{\mu,J}_n
&\= -\frac{n+1}{12 n} \frac{1}{\beta} \frac{2\ell_A \ell_B}{L^2} - 4 \sum_{m,l=1}^{\infty} \frac{1}{l } e^{-\frac{4\pi^2 l}{\beta} m} \cos [2\pi\mu l] \nonumber \\
& \times  \frac{1}{n\-1} \bigg[n-  \frac{\sinh \left(\! \frac{\pi}{\beta} \frac{l\ell_A}{L} \!\right)}{\sinh \left(\! \frac{\pi}{\beta} \frac{l}{n} \frac{\ell_A}{L} \!\right)} -  \frac{\sinh \left(\! \frac{\pi}{\beta}\frac{l\ell_B}{L} \!\right)}{\sinh \left(\! \frac{\pi}{\beta} \frac{l}{n} \frac{\ell_B}{L} \!\right)} +  \frac{\sinh \big(\! \frac{\pi}{\beta} \frac{l(\ell_A+\ell_B)}{L} \big)}{\sinh \left(\! \frac{\pi}{\beta} \frac{l}{n} \frac{\ell_A+\ell_B}{L} \!\right)} \bigg] \;. 
\end{align} 
Again we need to add \eqref{MutualInfoNoBGFHighTAlphaZero} to get the full result. 
For $\alpha \neq 0$,
\begin{align}\label{GeneralMutualInformationHighTPeriAplhaZero}
\tilde I^{\mu,J}_n 
&\=\frac{2}{1\- n} \sum_{m=1}^{\infty} \sum_{k=\frac{1-n}{2}}^{\frac{n-1}{2}} \!\! \log \! \Bigg[\! \frac{\big[\! \cos (2\pi [J \+ \frac{k}{n \alpha}\frac{\ell_A}{L}] ) \- \cos (\frac{4\pi^2}{\alpha} m)\big]\!
\big[\! \cos (2\pi [J \+ \frac{k}{n \alpha}\frac{\ell_B}{L}] ) \- \cos (\frac{4\pi^2}{\alpha} m))\big]}{\big[\cos (2\pi J ) \- \cos (\frac{4\pi^2}{\alpha} m )\big] 
\big[ \cos (2\pi [J + \frac{k}{n \alpha}\frac{\ell_A+\ell_B}{L}] ) \- \cos (\frac{4\pi^2}{\alpha} m)\big]} \Bigg] \;. 
\end{align} 
This also need to be combined to have the full result with \eqref{MutualInfoNoBGFHighTAlphaNonZero}. The mutual (R\'enyi) information is an oscillating function of the current $J$.

\section{Outlook}\label{sec:conclusion}

In this paper we carry out explicit and detailed computations of the entanglement entropy, the R\'enyi entropy and the mutual information of the 2-dimensional Dirac fermions in the presence of the background gauge fields. We summarize the salient features in the introduction and also previously in \cite{Kim:2017xoh}. Here we comments on some future directions. 

First of all, it would be interesting to perform similar computations for discrete models (for example, \cite{Vidal:2002rm}\cite{Latorre:2003kg}\cite{Korepin:2003}\cite{Eisert:2008ur}) with some background gauge fields for appropriate boundary conditions to check whether they would bear similar physical properties such as chemical potential and current dependences of the entropies at zero temperature, beat frequency, and more. 
Precise computations of the entropies and especially the mutual information emphasizing the dependence on the background fields and the boundary conditions would be helpful to identify them in available experiments \cite{Experiment}. 

In the presence of a current $J$, we have computed the entropies and mutual information for definite spin structures of the Dirac fermion by taking a periodic or anti-periodic boundary condition. These are the natural physical boundary conditions. It will be interesting to understand how the entropies interpolate these two different behaviors as a function of a parameter, the current $J$, that connects those two boundary conditions.  To do so, we need to reformulate the entropies and the mutual information for $0<a+J<1/2$. This turns out to require numerical approach that is beyond the scope of this paper. 

It is also interesting to study the role of the parameter $\alpha$ in the entropies. At the high temperature limit, there are two equivalent behaviors for the entropies depending on the presence or absence of the parameter. It is well known that $\alpha=0$ is connected to the usual thermodynamic entropy at high temperature. It is natural to ask whether the other limit with $\alpha\neq 0$ is physical or not. Apparently, there is no reason to consider that case as unphysical. Thus it is reasonable to investigate this case more carefully. Especially, one can ask how these two different limits are connected to the low temperature limit, where taking $\alpha\to 0$ or $\alpha \to \infty$ is a smooth limit. This is even more curious once we remind ourselves that the low and high temperature limits are equivalent because they are connected by modular transformation or Poisson resummation. Due to technical reasons, investigating this also requires a numerical method. 

\section*{Acknowledgments} 

We thank John Cardy, Paul de Lange, Mitsutoshi Fujita, Elias Kiritsis, Alfred Shapere, and especially Sumit Das  
for helpful discussions and numerous comments. 
This work is partially supported by NSF Grant PHY-1214341.  

\appendix

\section{A finite contribution for $\beta \to \infty$} \label{sec:AppendixFPEAZT}

We quote the answer for the sum over the index $k$ in the expression \eqref{RenyiFinite1}
\begin{align}
&\frac{1}{1\- n} \sum_{k=-\frac{n-1}{2}}^{\frac{n-1}{2}} \log \Big|\frac{(1 + y_1^{-1} q^{1/2})}{(1 + y_2^{-1} q^{1/2})} \Big|^2 
= \frac{1}{1\- n} \sum_{k=-\frac{n-1}{2}}^{\frac{n-1}{2}} \log \Big|\frac{(1 + e^{M - 2\pi i \frac{k}{n} \frac{\ell_t}{L}})}{(1 + e^{M})} \Big|^2 \nonumber \\
&=\frac{1}{1\- n} \log \left[\frac{e^{nM}}{(1+e^{nM})^{2n}} ~QPochhammer[-e^{M-\frac{\pi i(n-1)}{n} \frac{\ell_t}{L}},e^{\frac{2\pi i}{n} \frac{\ell_t}{L}},n] \right. \nonumber \\
&\hspace{1.6in} \times \left. QPochhammer[-e^{-M-\frac{\pi i(n-1)}{n} \frac{\ell_t}{L}},e^{\frac{2\pi i}{n} \frac{\ell_t}{L}},n] \right]
\;,
\end{align} 
where the $QPochhammer[a,q,n]$ symbol is given by 
\begin{align}
QPochhammer[a,q,n] = \left\{ \begin{array}{lll}
\prod_{j=0}^{n-1} (1- a q^j) \;,  & \qquad  & n>0 \;, \\ & & \\
1 \;,  & \qquad  & n=0 \;, \\ & & \\
\prod_{j=1}^{|n|} (1- a q^{-j})^{-1} \;,  & \qquad  & n<0 \;, \\  & & \\
\prod_{j=0}^{\infty} (1- a q^j) \;,  & \qquad  & n=\infty \;. \end{array} \right.
\end{align}


\begin{thebibliography}{99}

\bibitem{Holzhey:1994we} 
  C.~Holzhey, F.~Larsen and F.~Wilczek,
  ``Geometric and renormalized entropy in conformal field theory,''
  Nucl.\ Phys.\ B {\bf 424}, 443 (1994)
  doi:10.1016/0550-3213(94)90402-2
  [hep-th/9403108].
  
\bibitem{Calabrese:2004eu} 
  P.~Calabrese and J.~L.~Cardy,
  ``Entanglement entropy and quantum field theory,''
  J.\ Stat.\ Mech.\  {\bf 0406}, P06002 (2004)
  doi:10.1088/1742-5468/2004/06/P06002
  [hep-th/0405152].

\bibitem{Calabrese:2009qy} 
  P.~Calabrese and J.~Cardy,
  ``Entanglement entropy and conformal field theory,''
  J.\ Phys.\ A {\bf 42}, 504005 (2009)
  doi:10.1088/1751-8113/42/50/504005
  [arXiv:0905.4013 [cond-mat.stat-mech]].

\bibitem{Casini:2009sr} 
  H.~Casini and M.~Huerta,
  ``Entanglement entropy in free quantum field theory,''
  J.\ Phys.\ A {\bf 42}, 504007 (2009)
  doi:10.1088/1751-8113/42/50/504007
  [arXiv:0905.2562 [hep-th]].
  
\bibitem{Bennett:1995tk} 
  C.~H.~Bennett, H.~J.~Bernstein, S.~Popescu and B.~Schumacher,
  ``Concentrating partial entanglement by local operations,''
  Phys.\ Rev.\ A {\bf 53}, 2046 (1996)
  doi:10.1103/PhysRevA.53.2046
  [quant-ph/9511030].
   
\bibitem{Renyi} 
A. Renyi, ``On measures of entropy and information,'' in
Proc. Fourth Berkeley Symp. Math. Stat. Prob., 1960,
Vol. 1, 547, Berkeley, 1961. University of California Press.

\bibitem{Klebanov:2011uf} 
I.~R.~Klebanov, S.~S.~Pufu, S.~Sachdev and B.~R.~Safdi,
``Renyi Entropies for Free Field Theories,''
JHEP {\bf 1204}, 074 (2012)
doi:10.1007/JHEP04(2012)074
[arXiv:1111.6290 [hep-th]].

\bibitem{Belin:2013uta} 
A.~Belin, L.~Y.~Hung, A.~Maloney, S.~Matsuura, R.~C.~Myers and T.~Sierens,
``Holographic Charged Renyi Entropies,''
JHEP {\bf 1312}, 059 (2013)
doi:10.1007/JHEP12(2013)059
[arXiv:1310.4180 [hep-th]].

\bibitem{Ryu:2006bv} 
S.~Ryu and T.~Takayanagi,
``Holographic derivation of entanglement entropy from AdS/CFT,''
Phys.\ Rev.\ Lett.\  {\bf 96}, 181602 (2006)
doi:10.1103/PhysRevLett.96.181602
[hep-th/0603001].

\bibitem{Ryu:2006ef} 
S.~Ryu and T.~Takayanagi,
``Aspects of Holographic Entanglement Entropy,''
JHEP {\bf 0608}, 045 (2006)
doi:10.1088/1126-6708/2006/08/045
[hep-th/0605073].

\bibitem{Hubeny:2007xt} 
V.~E.~Hubeny, M.~Rangamani and T.~Takayanagi,
``A Covariant holographic entanglement entropy proposal,''
JHEP {\bf 0707}, 062 (2007)
doi:10.1088/1126-6708/2007/07/062
[arXiv:0705.0016 [hep-th]].

\bibitem{Nishioka:2009un} 
T.~Nishioka, S.~Ryu and T.~Takayanagi,
``Holographic Entanglement Entropy: An Overview,''
J.\ Phys.\ A {\bf 42}, 504008 (2009)
doi:10.1088/1751-8113/42/50/504008
[arXiv:0905.0932 [hep-th]].

\bibitem{Casini:2011kv} 
H.~Casini, M.~Huerta and R.~C.~Myers,
``Towards a derivation of holographic entanglement entropy,''
JHEP {\bf 1105}, 036 (2011)
doi:10.1007/JHEP05(2011)036
[arXiv:1102.0440 [hep-th]].

\bibitem{Lewkowycz:2013nqa} 
A.~Lewkowycz and J.~Maldacena,
``Generalized gravitational entropy,''
JHEP {\bf 1308}, 090 (2013)
doi:10.1007/JHEP08(2013)090
[arXiv:1304.4926 [hep-th]].

\bibitem{Ogawa:2011bz} 
  N.~Ogawa, T.~Takayanagi and T.~Ugajin,
  ``Holographic Fermi Surfaces and Entanglement Entropy,''
  JHEP {\bf 1201}, 125 (2012)
  doi:10.1007/JHEP01(2012)125
  [arXiv:1111.1023 [hep-th]].
  
\bibitem{Herzog:2013py} 
  C.~P.~Herzog and T.~Nishioka,
  ``Entanglement Entropy of a Massive Fermion on a Torus,''
  JHEP {\bf 1303}, 077 (2013)
  doi:10.1007/JHEP03(2013)077
  [arXiv:1301.0336 [hep-th]].

\bibitem{CardySlides:2016} 
  J.~Cardy,
  ``Entanglement in CFTs at Finite Chemical Potential,'' \href{http://www2.yukawa.kyoto-u.ac.jp/~entangle2016/YCardy.pdf}{presentation} at the Yukawa International Seminar 2016 (YKIS2016) ``Quantum Matter, Spacetime and Information.'' 

\bibitem{Arias:2014ksa} 
  R.~E.~Arias, D.~D.~Blanco and H.~Casini,
  ``Entanglement entropy as a witness of the Aharonov-Bohm effect in QFT,''
  J.\ Phys.\ A {\bf 48}, no. 14, 145401 (2015)
  doi:10.1088/1751-8113/48/14/145401
  [arXiv:1409.3269 [hep-th]].

\bibitem{Kim:2017xoh} 
  B.~S.~Kim,
  ``Entanglement Entropy, Current, and Chemical Potential,''
  arXiv:1705.01859 [hep-th].

\bibitem{DiFrancesco:1997nk} 
  P.~Di Francesco, P.~Mathieu and D.~Senechal,
  ``Conformal Field Theory,'' 1997, Springer
  doi:10.1007/978-1-4612-2256-9

\bibitem{Hori:2003ic} 
  K.~Hori, S.~Katz, A.~Klemm, R.~Pandharipande, R.~Thomas, C.~Vafa, R.~Vakil and E.~Zaslow,
  ``Mirror symmetry,'' 2003
  American Mathematical Society

\bibitem{Casini:2005rm} 
  H.~Casini, C.~D.~Fosco and M.~Huerta,
  ``Entanglement and alpha entropies for a massive Dirac field in two dimensions,''
  J.\ Stat.\ Mech.\  {\bf 0507}, P07007 (2005)
  doi:10.1088/1742-5468/2005/07/P07007
  [cond-mat/0505563].

\bibitem{Kabat:1995eq} 
  D.~N.~Kabat,
  ``Black hole entropy and entropy of entanglement,''
  Nucl.\ Phys.\ B {\bf 453}, 281 (1995)
  doi:10.1016/0550-3213(95)00443-V
  [hep-th/9503016].

\bibitem{Azeyanagi:2007bj} 
  T.~Azeyanagi, T.~Nishioka and T.~Takayanagi,
  ``Near Extremal Black Hole Entropy as Entanglement Entropy via AdS(2)/CFT(1),''
  Phys.\ Rev.\ D {\bf 77}, 064005 (2008)
  doi:10.1103/PhysRevD.77.064005
  [arXiv:0710.2956 [hep-th]].

\bibitem{Vidal:2002rm} 
G.~Vidal, J.~I.~Latorre, E.~Rico and A.~Kitaev,
``Entanglement in quantum critical phenomena,''
Phys.\ Rev.\ Lett.\  {\bf 90}, 227902 (2003)
doi:10.1103/PhysRevLett.90.227902
[quant-ph/0211074].

\bibitem{Latorre:2003kg} 
J.~I.~Latorre, E.~Rico and G.~Vidal,
``Ground state entanglement in quantum spin chains,''
Quant.\ Inf.\ Comput.\  {\bf 4}, 48 (2004)
[quant-ph/0304098].

\bibitem{Korepin:2003} 
V.~Korepin,
``Universality of Entropy Scaling in 1D Gap-less Models,''
Phys.\ Rev.\ Lett.\  {\bf 92}, 096402 (2005)
[cond-mat/0311056].

\bibitem{Eisert:2008ur} 
J.~Eisert, M.~Cramer and M.~B.~Plenio,
``Area laws for the entanglement entropy - a review,''
Rev.\ Mod.\ Phys.\  {\bf 82}, 277 (2010)
doi:10.1103/RevModPhys.82.277
[arXiv:0808.3773 [quant-ph]].

\bibitem{Experiment}
R. Islam,	R. Ma,	P. M. Preiss, M. E. Tai,	A. Lukin, M. Rispoli	and M. Greiner, 
``Measuring entanglement entropy in a quantum many-body system,''
Nature, {\bf 528}, 77 (2015).


\end{thebibliography}
\end{document}